\documentclass[10pt,journal,compsoc]{IEEEtran}
\usepackage[nocompress]{cite}

\usepackage{etoolbox}
\newtoggle{doublecolumn}
\newtoggle{calculateFigures}

\toggletrue{doublecolumn}
\togglefalse{calculateFigures}

\usepackage{etex}

\usepackage[english]{babel}
\usepackage{lipsum}
\usepackage{amsbsy,amsmath,amsfonts,amssymb,amsthm}
\usepackage{mathtools}
\usepackage{textcomp} 
\usepackage{relsize}

\usepackage{bm,cite,cases,pstricks,times,url,verbatim} 
\usepackage[noend]{algpseudocode}
\usepackage{booktabs}
\usepackage{tabularx,array,dcolumn,multirow}

\usepackage{algorithm}

\usepackage{yfonts}

\usepackage{xr} 

\usepackage{soul} 

\usepackage{blkarray,bigdelim} 

\allowdisplaybreaks[4]

\usepackage{centernot} 

\newcommand{\subparagraph}{}

\iftoggle{doublecolumn}{
    \newtheorem{thm}{Theorem}
    \newtheorem{fact}{Fact}
    \newtheorem{lemma}{Lemma}
    \newtheorem{definition}{Definition}
    \newtheorem{conj}{Conjecture}
    \newtheorem{propos}{Proposition}
    \newtheorem{corol}{Corollary}
    \newtheorem{ass}{Assumption}
    \newtheorem{example}{Example}
    \newtheorem{remark}{Remark}
    \newtheorem{note}{Note}
    \newtheorem{obs}{Observation}
}{

    \newtheoremstyle{exampstyle}
      {0} 
      {0} 
      {\itshape} 
      {} 
      {\bfseries} 
      {.} 
      {.5em} 
      {} 

    \theoremstyle{exampstyle} 
    \theoremstyle{exampstyle} 
    \theoremstyle{exampstyle} 
    \theoremstyle{exampstyle} 
    \theoremstyle{exampstyle} 
    \theoremstyle{exampstyle} 
    \theoremstyle{exampstyle} 
    \theoremstyle{exampstyle} \newtheorem{ass}{Assumption}
    \theoremstyle{exampstyle} 
    \theoremstyle{exampstyle} 
    \theoremstyle{exampstyle} 
    \theoremstyle{exampstyle} \newtheorem{obs}{Observation}
}

\newcommand{\argmax}[1]{\underset{#1}{\operatorname{arg}\,\operatorname{max}}\;}

\makeatletter
\newcommand{\pushright}[1]{\ifmeasuring@#1\else\omit\hfill$\displaystyle#1$\fi\ignorespaces}
\newcommand{\pushleft}[1]{\ifmeasuring@#1\else\omit$\displaystyle#1$\hfill\fi\ignorespaces}

\begingroup
\catcode`\#=11
\gdef\noautorotate{-dAutoRotatePages#/None}
\endgroup

\usepackage{graphicx,xcolor,float,dblfloatfix}
\usepackage{psfrag}

\usepackage{caption}
\usepackage{subcaption}

\newcommand{\subalign}[1]{%
  \vcenter{%
    \Let@ \restore@math@cr \default@tag
    \baselineskip\fontdimen10 \scriptfont\tw@
    \advance\baselineskip\fontdimen12 \scriptfont\tw@
    \lineskip\thr@@\fontdimen8 \scriptfont\thr@@
    \lineskiplimit\lineskip
    \ialign{\hfil$\m@th\textstyle##$&$\m@th\textstyle{}##$\crcr
      #1\crcr
    }%
  }
}

\iftoggle{calculateFigures}{
    \usepackage[crop=off,runs=2,pspdf=\noautorotate]{auto-pst-pdf}
}{
    \usepackage[off]{auto-pst-pdf}
}

\usepackage{hyperref}

\graphicspath{%
{./Figures/}
}

\usepackage{color}

\usepackage[compact]{titlesec}

\begin{document}

\author{Alessandro~Biason,~\IEEEmembership{Student~Member,~IEEE,} 
        Subhrakanti~Dey,~\IEEEmembership{Senior~Member,~IEEE,}  
        and~Michele~Zorzi,~\IEEEmembership{Fellow,~IEEE}
\IEEEcompsocitemizethanks{\IEEEcompsocthanksitem Alessandro~Biason and Michele~Zorzi are with the Department of Information Engineering, University of Padua, 35131 Padua, Italy.\protect\\
E-mail: biasonal@dei.unipd.it; zorzi@dei.unipd.it
\IEEEcompsocthanksitem Subhrakanti~Dey is with the Department of Engineering Science, Uppsala University, Uppsala, Sweden.\protect\\
E-mail: Subhrakanti.Dey@signal.uu.se
}
\thanks{A preliminary version of this paper has been accepted for presentation at IEEE WCNC 2017~\cite{Biason2017WCNC}.}
\thanks{This work was partially supported by Intel's Corporate Research Council.}
}

\title{A Decentralized Optimization Framework \\ for  Energy Harvesting Devices}

\thispagestyle{empty}


\IEEEtitleabstractindextext{%
\begin{abstract}
Designing decentralized policies for wireless communication networks is a crucial problem, which has only been partially solved in the literature so far. In this paper, we propose the Decentralized Markov Decision Process (Dec-MDP) framework to analyze a wireless sensor network with multiple users which access a common wireless channel. We consider devices with energy harvesting capabilities, so that they aim at balancing the energy arrivals with the data departures and with the probability of colliding with other nodes. Randomly over time, an access point triggers a SYNC slot, wherein it recomputes the optimal transmission parameters of the whole network, and distributes this information. Every node receives its own policy, which specifies how it should access the channel in the future, and, thereafter, proceeds in a \emph{fully decentralized} fashion, without interacting with other entities in the network. We propose a multi-layer Markov model, where an external MDP manages the jumps between SYNC slots, and an internal Dec-MDP computes the optimal policy in the near future. We numerically show that, because of the harvesting, a fully orthogonal scheme (e.g., TDMA-like) is suboptimal in energy harvesting scenarios, and the optimal trade-off lies between an orthogonal and a random access system.
\end{abstract}

\begin{IEEEkeywords}
admission control, energy consumption, energy harvesting, energy management, optimal scheduling, power control, telecommunication networks.
\end{IEEEkeywords}}

\maketitle

\IEEEdisplaynontitleabstractindextext
\IEEEpeerreviewmaketitle

\IEEEraisesectionheading{\section{Introduction}\label{sec:introduction}}

\IEEEPARstart{E}{nergy} Harvesting (EH) has been established as one of the most prominent solutions for prolonging the lifetime and enhancing the performance of Wireless Sensor Networks (WSNs) and Internet of Things (IoT) scenarios. Although this topic has been widely investigated in the literature so far, finding  proper energy management schemes is still an open issue in many cases of interest. In particular, using \emph{decentralized} policies, in which every node in the network acts autonomously and independently of the others, is a major problem of practical interest in WSNs where a central controller may not be used all the time. Many decentralized communication schemes (e.g., Aloha-like) can be found in the literature; however, most of them were designed without a principle of \emph{optimality}, i.e., without explicitly trying to maximize the network performance. Instead, in this work we characterize the optimal decentralized policy in a WSN with EH constraints and describe the related computational issues. Although this approach intrinsically leads to a more complex protocol definition, it also characterizes the maximum performance a network can achieve, and may serve as a baseline for defining quasi-optimal low-complexity protocols.

Energy harvesting related problems in WSNs have been addressed by several previous works (e.g., see~\cite{Ulukus2015} and the references therein), which tried to redefine many aspects of the communication devices, from both hardware and software perspectives. Indeed, EH devices need specialized equipment to harvest energy from the environment (e.g., a solar panel, or a rectenna) and to store it (e.g., miniaturized batteries). Clearly, the amount of harvested energy strongly depends on the EH source, therefore many different options have been proposed so far, including piezoelectric~\cite{Erturk2011}, temperature gradient~\cite{Stordeur1997}, daily temperature variation, vibrations~\cite{Beeby2006}, solar energy~\cite{Raghunathan2005}, indoor lights~\cite{Wang2008}, and Radio-Frequency (RF) energy~\cite{Lu2015}.

From a communication perspective, many analytical studies aimed at maximizing the performance of an energy harvesting network in terms of throughput~\cite{Michelusi2013,Biason2014,Tutuncuoglu2012a,Ozel2012b}, delay~\cite{Sharma2010}, quality of service~\cite{Pielli2016}, or other metrics. In particular, the problem of maximizing the average value of the reported data using a device with a rechargeable battery was formulated in~\cite{Lei2009}. An information theoretic analysis of an EH system can be found in~\cite{Ozel2012b}, where the authors presented the ``save-and-transmit'' and the ``best-effort-transmit'' schemes to achieve the channel capacity in the long run. Delay-aware strategies were presented in~\cite{Sharma2010}, where a single energy harvesting node is equipped with a data queue.
Some researchers focused on batteryless devices~\cite{Ozel2011,Smith2013}. In particular, \cite{Ozel2011}~considered a traditional EH system with amplitude constraints and found the channel capacity under causal channel information knowledge. Other kinds of batteryless devices are presented in~\cite{Smith2013}, in which a more innovative use of EH that exploits the RF waves as energy source is employed. Devices with finite batteries were studied in our previous works~\cite{Biason2015d,Biason2016b}, but also in~\cite{Tutuncuoglu2012a,Blasco2013}. A common technique to model the batteries is to approximate them with finite energy queues, in which energy arrives and departs over time. Markov models are suitable for these cases~\cite{Alsheikh2015} and were largely adopted in the literature so far~\cite{Lei2009,Blasco2013,Mohrehkesh2014,Michelusi2013,Biason2014,Biason2015d,Biason2016b}.
For more references about energy harvesting, we refer the readers to the surveys~\cite{Ulukus2015,Lu2015}.

However, differently from this paper, most of the protocols proposed in the literature considered isolated nodes and did not account for the interactions among devices, or focused on centralized policies, in which a controller coordinates all nodes and knows the global state of the system over time. \cite{Michelusi2015} analyzed decentralized policies with a particular focus on symmetric systems, and proposed a game theoretic approach for solving the problem. Instead, in this paper we use a different framework based on decentralized Markov decision processes, which can handle asymmetric scenarios. Decentralized theory was also used in~\cite{Hoang2014} for a wireless powered communication network. However, the scenario proposed therein is different from ours and the authors only focused on a narrow subclass of policies.

To model our energy harvesting system, we use the results about decentralized control theory recently developed by Dibangoye \emph{et al.}~\cite{Dibangoye2012,Dibangoye2013,Dibangoye2014}. In particular, \cite{Dibangoye2014} presented a detailed study of Decentralized-Partially Observable Markov Decision Processes (Dec-POMDPs) and proposed different approaches to solve them. The notion of \emph{occupancy state} was introduced as a fundamental building block for Dec-POMDPs, and it was shown that, differently from classic statistical descriptions (e.g., belief states), it represents a sufficient statistic for control purposes. Using the occupancy state, we can convert the Dec-POMDP in an equivalent MDP with a continuous state space, namely \emph{occupancy-MDP}. Then, standard techniques to solve POMDPs and MDPs can be applied; for example, an approach to solve a continuous state space MDP is to define a grid of points (see Lovejoy's grid approximation~\cite{Lovejoy1991}) and solve the MDP only in a subset of states. Although several papers introduced more advanced techniques to refine the grid~\cite{Zhou2001}, this approach may still be inefficient and difficult to apply. Instead, in this paper we use a different scheme, namely the Learning Real Time A$^*$ (LRTA$^*$) algorithm~\cite{Korf1990}, which has the key advantage of exploring only the states which are actually visited by the process, without the difficulty of defining a grid of points.

Converting the Dec-POMDP to an occupancy-MDP produces a simpler formulation of the problem, which however does not reduce its complexity. Indeed, for every occupancy state, it is still required to perform the \emph{exhaustive backup} operation, i.e., to compute a decentralized control policy. This is the most critical operation in decentralized optimization, since it involves solving a non-convex problem with many variables. Dibangoye \emph{et al.} proposed an alternate formulation of the exhaustive backup operation as a Constraint Program~\cite{Dechter2003}, which can be solved, e.g., using the bucket-elimination algorithm~\cite{Dechter1997}. The problem can be further simplified by imposing a predefined structure to the policy~\cite{Hoang2014}, so that only few parameters need to be optimized. While this may lead to suboptimal solutions, it greatly simplifies the numerical evaluation and, if correctly designed, produces close to optimal results. In our paper, we explore and compare both these directions.

\emph{\textbf{Problem Statement and Contributions.}} We consider a decentralized network with multiple devices and an access point that computes and distributes to all nodes the randomized transmission policy. A multi-layer Markov model, in which an internal Dec-MDP is nested inside an external MDP, is proposed and solved. The external layer models the time instants, namely SYNC slots, at which AP computes the policy, whereas the internal layer models the system evolution between consecutive SYNC slots. To solve the external layer, we use the Value Iteration Algorithm (see~\cite{Bertsekas2005}) as in~\cite{Blasco2013,Biason2016b}. However, differently from these papers, in our model the transition probabilities between states are derived from the optimization of the internal layer; moreover, the sojourn times in every state are not deterministic. Instead, the internal layer is solved using the Markov Policy Search algorithm~\cite{Dibangoye2012}. Because of the complexity of the optimal approach, we introduce two simpler schemes, which still exploit the structure of the optimal policy but can be computed in practice. In our numerical results we compare centralized and decentralized approaches, and discuss the performance loss of using a decentralized scheme.

Our main contributions can be summarized as follows
\begin{enumerate}
    \item We present a decentralized random access transmission scheme derived using a principle of \emph{optimality}, and discuss which are the computational pitfalls of this approach;
    \item We introduce two suboptimal policies, which are closely related to the structure of the optimal policy but can be numerically computed with reasonable complexity. These can be used as a baseline for developing heuristic schemes and real-time protocols. Moreover, although we present these approaches for an EH scenario, they may also be used in other contexts;
    \item We show that a decentralized scheme, if correctly designed, may achieve high performance, comparable with that of centralized solutions, while greatly reducing the signaling in the network;
    \item Finally, our most important contribution is to show that, differently from traditional networks (i.e., without energy constraints), where orthogonal resource allocation is optimal, the best transmission policy with energy harvesting is a hybrid approach between random and orthogonal access.
\end{enumerate}

The paper is organized as follows. Section~\ref{sec:system_model} presents the system model. The internal layer is described in Section~\ref{sec:internal}, whereas the external layer and the optimization problem are shown in Section~\ref{sec:ext_and_opt_prob}. Optimal and suboptimal solutions are derived in Sections~\ref{sec:opt_solution} and~\ref{sec:sub_opt_solutions}, respectively. The numerical results are shown in Section~\ref{sec:num_eval}. Finally, Section~\ref{sec:conclusions} concludes the paper.

\emph{\textbf{Notation.}} Throughout this paper, superscripts indicate the node indices, whereas subscripts are used for time indices. Boldface letters indicate \emph{global} quantities (i.e., vectors referred to all users), \emph{e.g.}, $\mathbf{e} \triangleq \langle e^1,\ldots,e^N\rangle$.

\begin{figure}[!t]
  \centering
  \includegraphics[trim = 0mm 0mm 0mm 0mm, width=1\columnwidth]{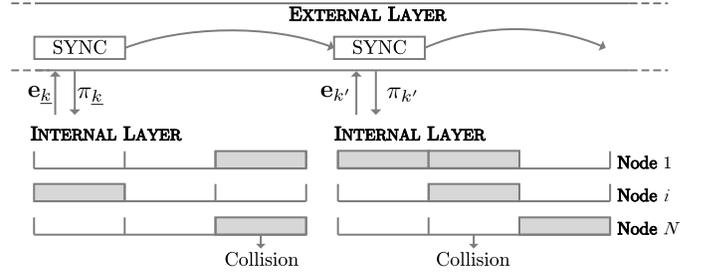}
  \caption{Time evolution of the system. After the SYNC slots every user acts independently of the others.}
  \label{fig:model}
\end{figure}

\section{System Model}\label{sec:system_model}

The network is composed of one Access Point (AP) and $N$ harvesting nodes (see \figurename~\ref{fig:model} for a graphical illustration).
We focus on an infinite time horizon framework, where a time slot $k$ corresponds to the time interval $[k\,\tau,(k+1)\tau)$, $k = 0,1,\ldots$, where $\tau$ is the common duration of all slots. During a slot, every node independently decides whether to access the uplink channel and transmit a message to AP, or to remain idle. We adopt an on/off collision model in which overlapping packet transmissions are always irrecoverable.

In slot $k$, node $i$ harvests energy from the environment according to a pdf $B_i^{(k)}$ (e.g., similarly to~\cite{Michelusi2012,Wang2015,Shaviv2015}, in this paper we will use a Bernoulli energy arrival process) and we assume independent arrivals among nodes. However, the model can be extended to the more general, correlated case (e.g., via an underlying common Markov model as in~\cite{Michelusi2013}).

Every node is equipped with a rechargeable battery, so that the energy stored in slot $k$ can be used in a later slot. 
The global energy level vector in slot $k$ is $\mathbf{e}_k = \langle e_k^1,\ldots,e_k^N \rangle$. Vector $\mathbf{e}_k$ is not known to the devices, which only see their own battery status, and is known to AP only in certain slots, namely ``SYNC slots''. In particular, in slot $k$, with probability (w.p.) $\beta_k \in (0,1]$, AP may trigger a SYNC slot and request all nodes to share their energy levels so that it can acquire $\mathbf{e}_k$.
We neglect the energy costs of these synchronization messages, which however, if considered, would make the benefits of using decentralized policies even higher.
For example, when $\beta_k = 1, \, \forall k$, AP has full knowledge of the battery levels at every time slot, and our model degenerates to~\cite{Biason2014}; instead, if $\beta_k = \beta_0, \, \forall k$, where $\beta_0$ is a constant value in $(0,1]$, AP uses a fully stochastic approach and asks for $\mathbf{e}_k$ with the same probability in every slot.

AP uses the information about $\mathbf{e}_k$ to initialize the transmission parameters of the whole network. Therefore, every time $\mathbf{e}_k$ is acquired (a SYNC slot), a coordination phase is performed and AP disseminates the \emph{policy} to all nodes (the policy is decentralized, so every node receives only its own policy).\footnote{We note that, although a user may also receive the policy of other devices, this information would not be useful. Indeed, the decentralized transmission policy is jointly designed by AP, therefore it implicitly considers the contributions of all nodes.} Thereafter, every device acts independently of the others until the next SYNC slot.

Although the proposed framework is very simple, modeling and solving it formally requires a complex mathematical structure. In particular, we decompose the system in two nested layers:
\begin{itemize}
    \item The external layer considers the jumps between consecutive SYNC slots. Indeed, since the global battery level $\mathbf{e}_k$ is completely known at every SYNC state, the system follows a Markov evolution;
    \item The internal layer models the actions to take between two SYNC slots and requires to compute a decentralized policy given $\mathbf{e}_k$. This will be modeled as a Dec-MDP, since multiple devices indirectly collaborate to achieve a common goal.
\end{itemize}

The two layers will be analyzed in the following sections.

\section{Internal Layer}\label{sec:internal}

We first consider the internal layer and present a mathematical tool to model the actions of the devices between two SYNC slots. In particular, we adopt the decentralized-Markov Decision Process (Dec-MDP) framework~\cite{Dibangoye2013}, which in our context is formally defined as follows. 

\subsection{Dececentralized--MDPs for EH Systems}\label{subsec:Dec_MDP}
An $N$-user Dec-MDP $\mathcal{M} = (\underline{k},\boldsymbol{\mathcal{E}},\boldsymbol{\mathcal{A}},p_{\rm int},r,\eta_0,\beta)$ is specified by~
\begin{itemize}
    \item \emph{\textbf{Initial Index.}} $\underline{k}$ represents the index of the SYNC slot that triggers the beginning of the internal layer. Thus, all the slots $k < \underline{k}$ are of no interest in this section. Note that, since $\beta_k$ is a probability, the position of the next SYNC slot is unknown a priori, therefore the time horizon of the Dec-MDP $\mathcal{M}$ begins at $\underline{k}$ and may extend to $+\infty$;
    
    \item \emph{\textbf{Battery Level.}} $\boldsymbol{\mathcal{E}} = \mathcal{E}^1 \times \cdots \times \mathcal{E}^N$ is the set of global battery levels $\mathbf{e}_k = \langle e_k^1,\ldots,e_k^N\rangle$, with $e_k^i \in \mathcal{E}^i \triangleq \{0,\ldots,e_{\rm max}^i\}$ (device $i$ can store up to $e_{\rm max}^i$ discrete energy quanta according to Equation~\eqref{eq:battery_evol}). Throughout, the terms ``battery level'' or ``state'' will be used interchangeably;
    
    \item \emph{\textbf{Action.}} $\boldsymbol{\mathcal{A}} = \mathcal{A}^1 \times \cdots \times \mathcal{A}^N$ is the set of global actions $\mathbf{a}_k = \langle a_k^1,\ldots,a_k^N\rangle$, where $a_k^i \in \mathcal{A}^i \triangleq [0,1]$ denotes the transmission probability. Although $a_k^i$ should assume continuous values, we only consider $S_a$ uniformly distributed samples of the interval $[0,1]$ for numerical tractability. Action $a_k^i$ is chosen by user $i$ in slot $k \geq \underline{k}$ throughout a function $\sigma_k^i: \mathcal{E}^i \to \mathcal{A}^i$, and depends only on the local state $e_k^i$. Finding $\sigma_k^i$ will be the objective of the optimization problem;
    
    \item \emph{\textbf{Transition Probability.}} $p_{\rm int}$ is the probability transition function $p_{\rm int}: \boldsymbol{\mathcal{E}} \times \boldsymbol{\mathcal{A}} \times \boldsymbol{\mathcal{E}} \to [0,1]$ which defines the probability $p_{\rm int}(\bar{\mathbf{e}}|\mathbf{e},\mathbf{a})$ of moving from a global battery level $\mathbf{e} = \langle e^1,\ldots,e^N \rangle \in \boldsymbol{\mathcal{E}}$ to a global battery level $\bar{\mathbf{e}} = \langle \bar{e}^1,\ldots,\bar{e}^N \rangle \in \boldsymbol{\mathcal{E}}$ under the global action $\mathbf{a} \in \boldsymbol{\mathcal{A}}$. When a transmission is performed, $m^i \geq 1$ energy quanta are consumed;\footnote{Quantity $m^i$ is associated with the physical parameters of the devices, and relates the amount of harvested energy with the energy used for transmission. Therefore, to transmit a single packet, in general, more than one energy quantum may be required.}
    
    \item \emph{\textbf{Reward.}} $r$ is the reward function $r: \boldsymbol{\mathcal{E}} \times \boldsymbol{\mathcal{A}} \to \mathbb{R}^+$ that maps the global action $\mathbf{a} \in \boldsymbol{\mathcal{A}}$ to the reward $r(\mathbf{e},\mathbf{a})$ when the global state is $\mathbf{e} \in \boldsymbol{\mathcal{E}}$;
    
    \item \emph{\textbf{Initial State Distribution.}} $\eta_{\underline{k}}$ is the initial state distribution. In our scenario we take~
    \begin{align}\label{eq:eta_0}
        \eta_{\underline{k}}(\mathbf{e}) = \begin{cases}
            1, \quad & \mbox{if } \mathbf{e} = \mathbf{e}_{\underline{k}}, \\
            0, \quad & \mbox{if } \mathbf{e} \in \boldsymbol{\mathcal{E}} \setminus \{\mathbf{e}_{\underline{k}}\},
        \end{cases}
    \end{align}
    
    \noindent where $\mathbf{e}_{\underline{k}}$ is the global state in correspondence of the initial SYNC slot and is fully known by AP;
    
    \item \emph{\textbf{SYNC Probability.}} $\beta$ represents the sequence $\beta_{\underline{k}},$ $\beta_{\underline{k}+1},\ldots$, which are the probabilities that a SYNC slot occurs. 
\end{itemize}

In Section~\ref{subsec:opt_problem} we will describe the optimization problem related to $\mathcal{M}$. Its solution provides a \emph{decentralized control policy}, which will be discussed in Sections~\ref{sec:opt_solution} and~\ref{sec:sub_opt_solutions}.

Before presenting in more detail the previous bullet points, it is important to emphasize the following key characteristics of the Dec-MDP under investigation:
\begin{itemize}
    \item $\mathcal{M}$ is \emph{jointly fully observable}, i.e., if all nodes collaborated and shared their local energy levels, the global state would be completely known (actually, this is what differentiates Dec-MDPs from Dec-POMDPs~\cite{Amato2013});
    \item $\mathcal{M}$ is a \emph{transition independent} Dec-MDP, i.e., the action taken by node $i$ influences only its own battery evolution in that slot and \emph{not} the others. Formally, the transition probability function $p_{\rm int}$ can be decomposed as~
    \begin{align}\label{eq:trans_indep}
        p_{\rm int}(\bar{\mathbf{e}}|\mathbf{e},\mathbf{a}) = \prod_{i = 1}^N p_{\rm int}^i(\bar{e}^i|e^i,a^i).
    \end{align}
    
    This feature is important to develop compact representations of the transmission policies, and in particular to derive Markovian policies as discussed in our Section~\ref{subsec:reward} and in~\cite[Theorem~1]{Dibangoye2012}.
\end{itemize}

\subsection{Battery Level}

We adopt a discrete model for the energy-related quantities, so that every battery can be referred to as an energy queue, in which arrivals coincide with the energy harvesting process, and departures with packet transmissions. In particular, the battery level of node $i$ in slot $k$ is $e_k^i$ and evolves as~
\begin{align}\label{eq:battery_evol}
    e_{k+1}^i = \min\{e_{\rm max}^i, e_k^i - s_k^i + b_k^i\},
\end{align}

\noindent where the $\min$ accounts for the finite battery size, $s_k^i$ is the energy used for transmission and $b_k^i$ is the energy arrived in slot $k$. $s_k^i$ is equal to $0$ w.p. $1-a_k^i$, and to $m^i$ w.p. $a_k^i$.

Note that this model has been widely used in the EH literature~\cite{Blasco2013,Michelusi2015,Wang2015}, and represents a good approximation of a real battery when $e_{\rm max}^i$ is sufficiently high.

\subsection{Action}

Node $i$ in slot $k$ can decide to access the channel w.p. $a_k^i = \sigma_k^i(e_k^i)$, or to remain idle w.p. $1-a_k^i$. When a transmission is performed, $m^i$ energy quanta are drained from the battery, and a corresponding reward $g(a_k^i)$ is obtained. When $e_k^i < m^i$, no transmission can be performed and $a_k^i = 0$.

\subsection{Transition Probability}

The transition probability function of user $i$, namely $p_{\rm int}^i$ (see Equation~\eqref{eq:trans_indep}), is defined as follows (for presentation simplicity, assume $\bar{e}^i < e_{\rm max}^i$ and $e^i \geq m^i$)~
\begin{align}
    \begin{split}
        &p_{\rm int}^i(\bar{e}^i|e^i,a^i) \\
        &= \begin{cases}
            (1-p_B^i) a^i, \quad & \mbox{if } \bar{e}^i = e^i - m^i, \\
            (1-p_B^i) (1-a^i) + p_B^i a^i, \quad & \mbox{if } \bar{e}^i = e^i - m^i+1, \\
            p_B^i (1-a^i), \quad & \mbox{if } \bar{e}^i = e^i + 1, \\
            0, \quad & \mbox{otherwise},
        \end{cases}
    \end{split}
\end{align}

\noindent where $p_B^i$ is the probability that user $i$ harvests one energy quantum. More sophisticated models, in which an arbitrary number of energy quanta can be simultaneously extracted, are described in~\cite{Biason2015d}, and can be integrated into our model (involving, however, higher computational costs).

\subsection{Reward} \label{subsec:reward}

We will use the term ``global reward'' to indicate the overall performance of the system in a slot, and simply ``single-user reward'' to refer to the performance of individual users.

\emph{\textbf{Single-User Reward.}} Assume to study isolated users, which do not suffer from interference, as in~\cite{Michelusi2013}. Data messages are associated with a \emph{potential reward}, described by a random variable $V^i$ which evolves independently over time and among nodes. The realization $\nu_k^i$ is perfectly known only at a time $t \geq k\,\tau$ and only to node $i$; for $t < k\,\tau$, only a statistical knowledge is available. Every node can decide to transmit (and accrue the potential reward $\nu_k^i$) or not in the current slot $k$ according to its value $\nu_k^i$.
In particular, it can be shown that a threshold transmission model is optimal for this system~\cite{Michelusi2013}; thus, node $i$ always transmits when $\nu_k^i \geq \nu_{\rm th}^i(e^i)$ and does not otherwise. Note that $\nu_{\rm th}^i(e^i)$ depends on the underlying state (battery level) of user $i$ but not on the time index $k$ (thus a \emph{stationary} scheduler can be developed).

On average, the reward of user $i$ in a single slot when the battery level is $e^i$ will be~
\begin{align}
    g(\nu_{\rm th}^i(e^i)) \triangleq \mathbb{E}[\chi(V^i \geq \nu_{\rm th}^i(e^i)) V^i] = \int_{\nu_{\rm th}^i(e^i)}^\infty \nu f_V^i(\nu) \ \mbox{d}\nu,
\end{align}

\noindent where $\chi(\cdot)$ is the indicator function and $f_V^i(\cdot)$ is the pdf of the potential reward, $V^i$.
It is now clear that the transmission probability $a^i$ is inherently dependent on the battery level as~
\begin{align}
    a^i = \sigma^i(e^i) = \int_{\nu_{\rm th}^i(e^i)}^\infty f_V^i(v) \ \mbox{d}v = \bar{F}_V^i(\nu_{\rm th}^i(e^i)),
\end{align}

\noindent where we explicitly introduced the function $\sigma^i(e^i)$, which maps local observations ($e^i$) to local actions $\sigma^i(e^i) = a^i$. Note that the complementary cumulative distribution function $\bar{F}_V^i(\cdot)$ is strictly decreasing and thus can be inverted. Therefore, there exists a one-to-one mapping between the threshold values and the transmission probabilities. In the following, we will always deal with $a^i$ instead of $\nu_{\rm th}^i(\cdot)$, and write $g(a^i)$ with a slight abuse of notation.

It can be proved that $g(a^i)$ is increasing and concave in $a^i$, i.e., transmitting more often leads to higher rewards, but with diminishing returns. Finally, note that this model is quite general and, depending on the meaning of $V^i$, can be adapted to different scenarios. For example, in a standard communication system in which the goal is the throughput maximization, $V^i$ can be interpreted as the transmission rate subject to fading fluctuations~\cite{Michelusi2012}.

\emph{\textbf{Global Reward.}}
The global reward is zero when multiple nodes transmit simultaneously, whereas it is equal to $w^i \nu_k^i$ if only node $i$ transmits in slot $k$ ($w^i$ is the weight of node $i$). On average, since the potential rewards are independent among nodes, we have~
\begin{align}
    \begin{split}
        r(\nu_{{\rm th},k}(\mathbf{e}_k)) =\ & \mathbb{E}\bigg[\sum_{i = 1}^N w^i \, V_k^i \, \chi(V_k^i \geq \nu_{{\rm th},k}^i(e_k^i))\\
        & \times \prod_{j\neq i} \chi(V_k^j < \nu_{{\rm th},k}^j(e_k^j)) \bigg],
    \end{split}
\end{align}
\noindent which can be rewritten as~
\begin{align}\label{eq:r_a}
    r(\mathbf{a}_k) = r(\boldsymbol{\sigma}_k(\mathbf{e}_k)) = \sum_{i = 1}^N w^i g(a_k^i) \prod_{j \neq i} (1-a_k^j),
\end{align}

\noindent where we used $\mathbf{a}_k$ instead of $\nu_{{\rm th},k}(\mathbf{e}_k)$ for ease of notation, and we introduced the vector function $\boldsymbol{\sigma}_k \triangleq \langle \sigma_k^1,\ldots,\sigma_k^N \rangle$.  We highlight that $\boldsymbol{\sigma}_k$ summarizes the actions of all users given every battery level in slot $k$, i.e., it specifies all the following quantities~
\begin{align}\label{eq:sigma_struct}
    \begin{matrix}
        \sigma_k^1(0) & \ldots & \sigma_k^1(e_{\rm max}^1),\\
        \vdots \\
        \sigma_k^N(0) & \ldots & \sigma_k^N(e_{\rm max}^N).
    \end{matrix}
\end{align}

\noindent Finding $\boldsymbol{\sigma}_{\underline{k}}, \boldsymbol{\sigma}_{\underline{k}+1}, \ldots$ represents the biggest challenge when solving a Dec-MDP.

\begin{figure}[!t]
  \centering
  \includegraphics[trim = 0mm 0mm 0mm 0mm, width=1\columnwidth]{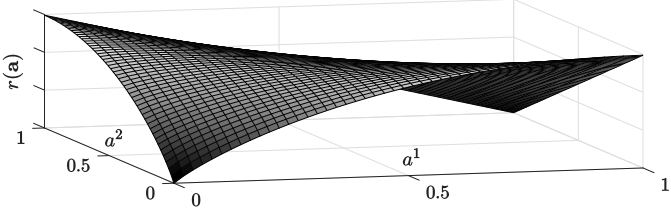}
  \caption{Global reward $r(\mathbf{a})$ when $N = 2$.}
  \label{fig:r_a}
\end{figure}

An important observation is that the reward~\eqref{eq:r_a} is not necessarily increasing nor convex in $\mathbf{a}$, which significantly complicates the solution. An example of $r(\mathbf{a})$ for the two-user case can be seen in \figurename~\ref{fig:r_a}. Note that the maximum is achieved when only one device transmits w.p. $1$ and the other does not transmit. 
This implies that, when the devices are not energy constrained (i.e., they have enough energy for transmitting and the current transmission policy does not influence the future), the optimal user allocation should follow an orthogonal approach so as to avoid collisions (the corner points $\langle a^1,a^2 \rangle = \langle 1,0 \rangle$ and $\langle a^1,a^2 \rangle = \langle 0,1 \rangle$ achieve the maximum reward). However, as we will discuss later, this observation does not hold in EH scenarios, in which an action in the current slot influences the future energy levels and, consequently, the future rewards.

Note that, in the previous expressions, we have implicitly restricted our study to Markovian policies, which map local observations to local actions (i.e., $\boldsymbol{\sigma}_k(\mathbf{e}_k) = \mathbf{a}_k$). In general decentralized frameworks, tracking previous observations can be used to optimally decide the current action (i.e., $\mathbf{a}_k$ depends on $\mathbf{e}_{\underline{k}},\ldots,\mathbf{e}_k$). However, it can be proved~\cite{Dibangoye2012} that under transition independent conditions (which hold in our case, see Section~\ref{subsec:Dec_MDP}), Markovian policies are optimal and thus keeping track of previous states is not necessary.

Finally, we note that using a stationary scheduler is suboptimal in the multi-user case (this will be very clear from our numerical evaluation, e.g., see~\figurename~\ref{fig:txProb_Pe1}); that is, even if we had $e_{k^\prime}^i = e_{k^{\prime\prime}}^i$ for $k^\prime \neq k^{\prime\prime}$ (same energy level in two different slots), using $\sigma_{k^\prime}^i(e_{k^\prime}^i) = \sigma_{k^{\prime\prime}}^i(e_{k^{\prime\prime}}^i)$ (same policy in two different slots) would \emph{not} be optimal. Unfortunately, this implies that the decentralized optimization is much more challenging than the centralized one, since the optimal policy should be computed for every slot $k \geq \underline{k}$.\footnote{In theory, the policy should be computed for an infinite number of slots. However, as will be explained in Equation~\eqref{eq:max_long}, when $k \gg \underline{k}$ the rewards obtained will be very small; therefore, it will not be necessary to compute the policy optimally for every $k \geq \underline{k}$.}

\subsection{Occupancy State} \label{subsec:occupancy_state}

Before formulating the optimization problem in the next section, we first introduce the concept of \emph{occupancy state}.

The occupancy state $\eta_k$ is defined as~
\begin{align}\label{eq:eta_k_def}
    \eta_k(\bar{\mathbf{e}}) \triangleq \mathbb{P}(\mathbf{e}_k = \bar{\mathbf{e}} | \eta_{\underline{k}},\boldsymbol{\sigma}_{\underline{k}},\ldots,\boldsymbol{\sigma}_{k-1}), \quad k \geq \underline{k}
\end{align}

\noindent and represents a probability distribution over the battery levels given the initial distribution $\eta_{\underline{k}}$ (introduced in Equation~\eqref{eq:eta_0}) and all decentralized decision rules prior to $k$.

It can be shown that the occupancy state represents a sufficient statistic for control purposes in Dec-MDPs,\footnote{Intuitively, the occupancy state replaces the state of the system in centralized MDPs, or the belief in POMDPs.} and can be easily updated at every slot using old occupancy states~($k \!>\! \underline{k}$):~
\begin{align}\label{eq:eta_k_update}
    \eta_k(\bar{\mathbf{e}}) = \omega(\eta_{k-1},\boldsymbol{\sigma}_{k-1}) = \sum_{\mathbf{e}} p(\bar{\mathbf{e}} | \mathbf{e}, \boldsymbol{\sigma}_{k-1}(\mathbf{e})) \eta_{k-1}(\mathbf{e}),
\end{align}

\noindent where $\omega$ is the occupancy update function.

\emph{\textbf{Occupancy-MDP.}}
Dibangoye \emph{et al.}~\cite{Dibangoye2014} developed a technique to solve Dec-MDPs by recasting them in equivalent continuous state MDPs. Similarly to the reduction techniques of POMDPs, in which the belief is used as the state in an equivalent MDP, for Dec-MDPs the occupancy state will represent the building block of the equivalent MDP (called \emph{occupancy-MDP}). In particular, the state space of the occupancy-MDP is the occupancy simplex, the transition rule is given by~\eqref{eq:eta_k_update}, the action space is $\boldsymbol{\mathcal{A}}$, and the instantaneous reward corresponding to decentralized decision rule $\boldsymbol{\sigma}_k$ is~
\begin{align}\label{eq:rho_def}
    \rho(\eta_k,\boldsymbol{\sigma}_k) = \sum_{\bar{\mathbf{e}} \in \boldsymbol{\mathcal{E}}} \eta_k(\bar{\mathbf{e}}) r(\boldsymbol{\sigma}_k(\bar{\mathbf{e}})),
\end{align}

\noindent Note that $\rho(\eta_k,\boldsymbol{\sigma}_k) \leq \max_{\bar{\mathbf{e}}} r(\boldsymbol{\sigma}_k(\bar{\mathbf{e}}))$, i.e., the loss of information corresponds to a lower reward. Moreover, note that if $\underline{k}$ were a SYNC slot, we would have $\rho(\eta_{\underline{k}},\boldsymbol{\sigma}_{\underline{k}}) = r(\boldsymbol{\sigma}_{\underline{k}}(\mathbf{e}_{\underline{k}}))$.

The complete structure of the occupancy-MDP will be given in Section~\ref{subsec:Bellman}.

\section{External Layer and Optimization Problem} \label{sec:ext_and_opt_prob}

So far, we have described how the system evolves between two SYNC slots. We now introduce the \emph{external Markov Chain}, which models the long-term evolution of the network by considering the subset of all slots composed only by the SYNC slots.

Assume that, without loss of generality, the first SYNC slot occurs at $\underline{k} = 0$, and that the state of the system is $\mathbf{e}_0$. According to Section~\ref{sec:internal}, AP uses $\mathbf{e}_0$ to compute and distribute to all nodes a decentralized policy $\boldsymbol{\sigma}_0,\boldsymbol{\sigma}_1,\ldots$. 
Moreover, the initial occupancy state $\eta_0$ is defined in~\eqref{eq:eta_0}, whereas the occupancy states $\eta_1,\eta_2,\ldots$ are evaluated as in Section~\ref{subsec:occupancy_state}. 

Assume now that the first SYNC slot after $\underline{k} = 0$ is slot $k^\prime$; thus, when the new SYNC slot occurs, we know that the transition probability from the initial global state $\mathbf{e}_0$ to the final global state $\bar{\mathbf{e}}$ is $\eta_{k^\prime}(\bar{\mathbf{e}})$ (i.e., it is given by the occupancy state by its definition), with $\eta_{\underline{k}} = \eta_0 = \chi\{\mathbf{e} = \mathbf{e}_0\}$.

Using $\eta_{k^\prime}(\bar{\mathbf{e}})$, we can compute the probability of going from $\mathbf{e}_0$ to $\bar{\mathbf{e}}$. Since $k^\prime$ is a random quantity ($\beta$ represents a sequence of probabilities), we provide the expression of the probability averaged over $k^\prime$ (this models the jumps of the external MC between SYNC states)~
\begin{align} \label{eq:p_e0_bar_e}
    p_{\rm ext}(\bar{\mathbf{e}} | \mathbf{e}_0) = \sum_{{k^\prime} = 1}^\infty \beta_{k^\prime} \bigg(\prod_{k^{\prime\prime} = 1}^{k^\prime-1} (1-\beta_{k^{\prime\prime}})\bigg) \eta_{k^\prime}(\bar{\mathbf{e}}).
\end{align}
    
\noindent In the previous expression, $k^\prime$ represents the index $>0$ of the first SYNC slot; $\beta_{k^\prime}$ is the probability that $k^\prime$ is a SYNC slot; the product $\prod_{k^{\prime\prime} = 1}^{k^\prime-1} (1-\beta_{k^{\prime\prime}})$ is the probability that no slots prior to $k^\prime$ are SYNC slots; $\eta_{k^\prime}(\bar{\mathbf{e}})$ is the transition probability, which implicitly depends on $\mathbf{e}_0$. We remark that, to evaluate~\eqref{eq:p_e0_bar_e}, we need the sequence $\boldsymbol{\sigma}_0, \boldsymbol{\sigma}_1, \ldots$ in order to compute all the future occupancy states $\eta_1,\eta_2,\ldots$. In Section~\ref{subsec:opt_problem}, we will specify how to define $\boldsymbol{\sigma}_0, \boldsymbol{\sigma}_1, \ldots$.

\begin{obs}
    The sequence of SYNC slots satisfies the Markov property, i.e., if $k^\prime$ is a SYNC slot, the system evolution for $k \geq k^\prime$ is conditionally independent of past states given $\mathbf{e}_{k^\prime}$. 
\end{obs}

We now formally define the optimization problem and link the internal and external layers.

\subsection{Optimization Problem} \label{subsec:opt_problem}

Define $\rho_{k|\underline{k}} \triangleq \rho(\eta_k,\boldsymbol{\sigma}_k)$ as the decentralized reward given the SYNC slot $\underline{k}$. Then, prior to the next SYNC slot, the reward of the system will be: $\rho_{\underline{k}|\underline{k}}$ w.p. $\beta_{\underline{k}+1}$, $\rho_{\underline{k}|\underline{k}} + \rho_{\underline{k}+1|\underline{k}}$ w.p. $(1-\beta_{\underline{k}+1})\beta_{\underline{k}+2}$, $\rho_{\underline{k}|\underline{k}} + \rho_{\underline{k}+1|\underline{k}} + \rho_{\underline{k}+2|\underline{k}}$ w.p. $(1-\beta_{\underline{k}+1})(1-\beta_{\underline{k}+2})\beta_{\underline{k}+3}$, and so forth. 
Summing together the previous terms, and taking the average over the energy harvesting processes, we obtain the normalized average reward~
\begin{align}
\label{eq:R_k_avg}
    R_{\underline{k}} \triangleq \ & \mathbb{E}\bigg[\sum_{k = \underline{k}}^\infty \rho_{k|\underline{k}} \, \sum_{k^\prime = k+1}^\infty \beta_{k^\prime} \prod_{k^{\prime\prime} = \underline{k}+1}^{k^\prime-1} (1-\beta_{k^{\prime\prime}}) \bigg].
\end{align}

The final goal of the system is to maximize the cumulative weighted undiscounted long-term reward,\footnote{We consider the undiscounted throughput in the long run because WSNs typically reach the steady-state conditions. The weights are used for fairness.} defined as~
\begin{align}\label{eq:G_beta_cent}
    G (\Pi,\mathbf{e}_0) = \liminf_{K \to \infty} \frac{1}{K} \sum_{\underline{k} = 0}^{K-1} \beta_{\underline{k}} \cdot R_{\underline{k}}(\pi_{\underline{k}},\mathbf{e}_{\underline{k}}).
\end{align}

\noindent $R_{\underline{k}}(\pi_{\underline{k}},\mathbf{e}_{\underline{k}})$ is given in~\eqref{eq:R_k_avg} when the initial state of the system is $\mathbf{e}_{\underline{k}}$ and a policy $\pi_{\underline{k}} \triangleq (\boldsymbol{\sigma}_{\underline{k}}^{\pi_{\underline{k}}},\boldsymbol{\sigma}_{\underline{k}+1}^{\pi_{\underline{k}}},\ldots)$ is employed. Policy $\pi_{\underline{k}}$ is decentralized, and it is drawn from $\Pi$, which includes all the decentralized transmission policies $\pi_{\underline{k}},\pi_{\underline{k}+1},\ldots$.

Since the sequence $\beta$ is a design parameter and its choice is arbitrary, we restrict our attention to the following case.
\begin{ass} \label{ass:beta_periodic}
    The SYNC probability sequence is periodic with period $q$ (i.e., $\beta_k = \beta_{k+q},\ \forall k$).
\end{ass}

\noindent For example, the simplest case is $q = 1$, and $\beta_k = \beta_0$ for every $k$. 
Under Assumption~\ref{ass:beta_periodic}, it can be shown that~\eqref{eq:G_beta_cent} is equivalent to~
\begin{align}\label{eq:G_steady}
    G (\Pi) = \sum_{\underline{k} = 0}^{q-1} \, \sum_{\mathbf{e} \in \boldsymbol{\mathcal{E}}}  \beta_{\underline{k}} \cdot R_{\underline{k}}(\pi_{\mathbf{e}},\mathbf{e}) \cdot {\rm ssp}_{\underline{k}}(\mathbf{e}),
\end{align}

\noindent where ${\rm ssp}_{\underline{k}}(\mathbf{e})$ is the steady-state probability of the global energy level $\mathbf{e}$ associated with $\beta_{\underline{k}}$, and, instead of iterating over all $k$, we take the sum over the energy levels (i.e., we iterate over the states of the external MC). Note that, in this case, the long-term undiscounted reward does not depend on the initial state of the system, therefore $G(\Pi,\mathbf{e}_0) = G(\Pi)$ for every $\mathbf{e}_0$.
The optimal solution of the external problem will be~
\begin{align} \label{eq:Pi_star}
    \Pi^\star = \argmax{\Pi} G (\Pi),
\end{align}

\noindent which is a Markov Decision Process (MDP). The underlying MC states are all the elements of $\boldsymbol{\mathcal{E}}$, whereas the actions, which influence the transition probabilities (Equation~\eqref{eq:p_e0_bar_e}), are given by the evolution of the internal Dec-MDP. In the following, for the sake of presentation simplicity, we impose $\beta_k = \beta_0,\ \forall k$ (i.e., $q = 1$). However, the results can be straightforwardly extended to the more general case.

\emph{\textbf{Value Iteration Algorithm (VIA).}} The optimization problem of Equation~\eqref{eq:Pi_star} can be solved using VIA~\cite[Vol.~1, Sec.~7.4]{Bertsekas2005}. Since we focus on $q = 1$, thanks to Equation~\eqref{eq:G_steady} we only examine $\underline{k} = 0$ (in the more general case $q > 1$, the procedure is analogous but with $q$ different equations). The Bellman equation~(see~\cite{Bertsekas2005}) to iteratively solve is~
\begin{align}\label{eq:Bellman_external}
    &z^{\rm iter}(\mathbf{e}) \gets \max_{\pi_{\mathbf{e}}} \Big\{\beta_0 \, R_0(\pi_{\mathbf{e}},\mathbf{e}) + \sum_{\bar{\mathbf{e}} \in \boldsymbol{\mathcal{E}}} p_{\rm ext}(\bar{\mathbf{e}} | \mathbf{e}) z^{{\rm iter}-1}(\bar{\mathbf{e}}) \Big\},
\end{align}

\noindent where ``iter'' is the index of VIA; $\beta_0 \, R_0(\pi_{\mathbf{e}},\mathbf{e})$ represents the initial reward, whereas the other term is the expected future reward (this is derived from the old values of the Bellman equation, $z^{{\rm iter}-1}(\cdot)$).
After a number of iterations (typically only a few), VIA converges, and the Bellman equations yield the optimal solution $G(\Pi^\star)$.

\section{Optimal Solution of the Internal Layer}\label{sec:opt_solution}

In the previous section we discussed the external optimization problem and its solution via the value iteration algorithm. However, every iteration of VIA requires to solve the $\max$ in Equation~\eqref{eq:Bellman_external}. This is equivalent to solving the Dec-MDP of the internal layer, since~\eqref{eq:Bellman_external} depends on the decentralized policy sequence $\pi_{\mathbf{e}}$.
In this section, we discuss how to do that optimally, whereas in Section~\ref{sec:sub_opt_solutions} we discuss suboptimal solutions.

\subsection{Bellman Equation} \label{subsec:Bellman}

The Bellman Equation~\eqref{eq:Bellman_external} can be rewritten using the definitions of $p_{\rm ext}(\bar{\mathbf{e}} | \mathbf{e}_0)$ (Equation~\eqref{eq:p_e0_bar_e}) and $R_0(\cdot)$ (Equation~\eqref{eq:R_k_avg}):
~
\begin{align} \label{eq:max_long}
    & \max_{\pi_{\mathbf{e}}} \bigg\{\beta_0 \, \mathbb{E}\bigg[\sum_{k = 0}^T \phi_k(\eta_k,\boldsymbol{\sigma}_k^{\pi_{\mathbf{e}}}) \Big| \mathbf{e} \bigg] \bigg\}, \\
    & \phi_k(\eta_k,\boldsymbol{\sigma}) \triangleq (1-\beta_0)^k \Big(\rho(\eta_k,\boldsymbol{\sigma}) + \sum_{\bar{\mathbf{e}} \in \boldsymbol{\mathcal{E}}} \eta_{k+1}(\bar{\mathbf{e}}) z^{{\rm iter}-1}(\bar{\mathbf{e}})\Big). \label{eq:phi_def}
\end{align}

\noindent Equation~\eqref{eq:max_long} represents the occupancy-MDP under investigation. For ease of notation, we used $\eta_{k+1} \triangleq \omega(\eta_k,\boldsymbol{\sigma})$. Note that $(1-\beta_0)^k$ decreases with $k$, whereas all the other terms are bounded. Consequently, $\phi_k(\cdot)$ decreases with $k$; thus, for large $k$, its contribution will be negligible. Because of this, we approximated the infinite sum with a finite sum from $0$ to $T$, where $T$ is a sufficiently large natural number. We now specify how the system behaves between $0$ and $T$, whereas the policy for $k > T$ can be arbitrarily chosen without degrading the performance of the system. 

We can solve the $\max$ by rewriting it in a recursive form:~
\begin{align}\label{eq:bar_v_def}
    v_k(\eta_k) = \max_{\boldsymbol{\sigma}} \begin{cases}
        \phi_k(\eta_k,\boldsymbol{\sigma}) + v_{k+1} (\omega(\eta_k,\boldsymbol{\sigma})), \ &\mbox{if } k < T, \\
        \phi_T(\eta_T,\boldsymbol{\sigma}), \ &\mbox{if } k = T,
    \end{cases}
\end{align}

\noindent where $v_k(\cdot)$ is the \emph{cost-to-go function}. Equation~\eqref{eq:max_long} is equivalent to $\beta_0\cdot v_0(\eta_0)$. The trivial solution to find $\pi_{\mathbf{e}}$ is to apply VIA in the finite horizon; however, this would require, for every $k$, to specify $v_k(\eta_k)$ for \emph{every} $\eta_k$, which is impossible in practice. 

An alternate solution is to use techniques originally developed for POMDPs which were later used for Dec-POMDPs. In particular, the Learning Real Time A$^*$ (LRTA$^*$) algorithm is suitable for our case, since it explores only the occupancy states which are actually visited during the planning horizon and avoids grid-based approaches (e.g., as used in~\cite{Lovejoy1991}). In~\cite{Dibangoye2012}, the Markov Policy Search (MPS) algorithm was introduced as an adaptation of LRTA$^*$ to decentralized scenarios. 

In summary, MPS operates as follows~
\begin{enumerate}
    \item It starts at $\underline{k} = 0$ and, for every $k \geq 0$, it computes the LHS of~\eqref{eq:bar_v_def} with LRTA$^*$, i.e., the maximization problem is solved only for the occupancy states which are actually visited and not for every $\eta_k$;
    \item It replaces $v_{k+1}$ in the RHS with an upper bound, which can be computed using the convexity of the cost-to-go function. In Section~\ref{subsec:ub} we will further discuss this point;
    \item When $k = T$ is reached, a lower bound of the optimal cost-to-go function is evaluated in a backward direction (see~\cite[Sec.~5.1]{Dibangoye2012}).
\end{enumerate}

The procedure is repeated until upper and lower bounds converge to the optimal solution. We refer the readers to~\cite{Dibangoye2012,Dibangoye2013} for a full description of the algorithm. In the following, we discuss how to find the upper bound of the cost-to-go function, which will be used as a building block in Section~\ref{subsec:CP}.

\subsection{Upper Bound of the Cost-to-go Function} \label{subsec:ub}

It can be shown by induction that the optimal cost-to-go function $v_k^\star$ is a convex function of the occupancy states and can be approximated by piecewise linear functions~\cite[Theorem~4.2]{Dibangoye2013}. The upper bound $\bar{v}_k$ of $v_k^\star$ can be written as~
\begin{align}    
    \bar{v}_k(\eta_k)& = \max_{\boldsymbol{\sigma}} \{ \phi_k(\eta_k,\boldsymbol{\sigma}) + C(\Upsilon_k,\omega(\eta_k,\boldsymbol{\sigma}))\}, \label{eq:v_beta_C}
\end{align}

\noindent where $C$ interpolates the occupancy state $\omega(\eta_k,\sigma)$ using the point set $\Upsilon_k$, which contains the visited occupancy states along with their upper bound values. Every time \eqref{eq:v_beta_C} is solved, a new point $(\eta_k,\bar{v}_k(\eta_k))$ is added to $\Upsilon_k$.
The first points to be put in $\Upsilon_k$ are the corners of the occupancy simplex (i.e., the $|\boldsymbol{\mathcal{E}}|$ points $[1,0,\ldots,0],\ldots,[0,\ldots,0,1]$) with their upper bound values obtained solving the following full knowledge MDP:~
\begin{align}\label{eq:R_beta_cent}
    \overline{R}_{0}(\pi_{\mathbf{e}},\mathbf{e})  \triangleq \mathbb{E}\bigg[ \sum_{k =  0}^\infty r(\boldsymbol{\sigma}_k^{\pi_{\mathbf{e}}}(\mathbf{e}_{k})) \, (1-\beta_0)^k \Big| \mathbf{e} \bigg],
\end{align}

\noindent which is equivalent to~\eqref{eq:R_k_avg} but with $r(\cdot)$ instead of $\rho_{k|0}$ and with $q = 1$. Expression~\eqref{eq:R_beta_cent} implicitly assumes that the state of the system $\mathbf{e}_k$ is globally known in slot $k$, i.e., a \emph{centralized}-oriented network. Since this is a standard MDP, it can be easily solved with VIA.

\begin{figure}[!t]
  \centering
  \includegraphics[trim = 0mm 0mm 0mm 0mm, width=.9\columnwidth]{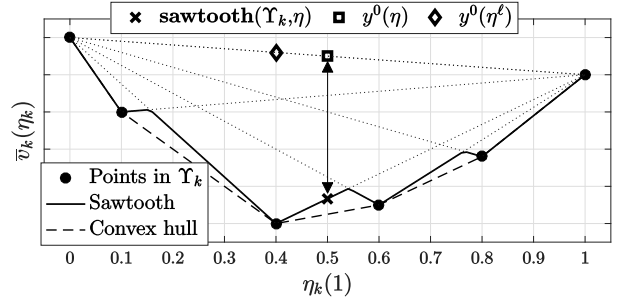}
  \caption{Sawtooth projection in the two-dimensional case. The arrow identifies the quantity $\max_{\ell \in \mathcal{L}} \{(y^0(\eta^\ell) \!-\! v^\ell) \xi^\ell \}$ when $\eta = \langle 0.5,0.5 \rangle$.}
  \label{fig:sawtooth}
\end{figure}

\emph{\textbf{Sawtooth Projection.}}
Ideally, we could use a linear interpolation as the function $C$ (i.e., map $\eta_k$ on the convex hull of point set $\Upsilon_k$), but this would incur high complexity. A faster approach, which however has shown good performance in many applications, is to replace $C$ with the sawtooth projection:\footnote{The term ``sawtooth'' comes from the shape of the interpolating function in the two-dimensional case (see \figurename~\ref{fig:sawtooth}). The idea of the approach is to interpolate a point $\eta$ using $|\boldsymbol{\mathcal{E}}|-1$ corner points of the simplex, and one point taken from $\Upsilon_k$ ($\ell$ in Equation~\eqref{eq:sawtooth_def}).}~
\begin{align} \label{eq:sawtooth_def}
    \textbf{swt}(\Upsilon_k,\eta) = \! y^0(\eta) - \max_{\ell \in \mathcal{L}} \{(y^0(\eta^\ell) - v^\ell) \xi^\ell \},
\end{align}

\noindent where $\eta$ is the occupancy state to interpolate, $(\eta^\ell,v^\ell)$ is the $\ell$-th element of $\Upsilon_k$, $\mathcal{L}$ is the set of indices of $\Upsilon_k$, $\xi^\ell$ is the interpolation coefficient, and $y^0(\cdot)$ is the upper bound computed using the corner points of $\Upsilon_k$, i.e.,
\begin{align}\label{eq:y_0_def}
    y^0(\eta) = \sum_{\mathbf{e} \in \boldsymbol{\mathcal{E}}} \eta(\mathbf{e}) \Upsilon_k(\mathbf{e}).
\end{align}

\noindent In the previous expression, with a slight abuse of notation, $\Upsilon_k(\mathbf{e})$ indicates the upper bound value at the corner $\mathbf{e}$ of the simplex. The interpolation coefficient is defined as~
\begin{align}
    \xi^\ell \triangleq \min_{\mathbf{e}\, :\, \eta^\ell(\mathbf{e})>0} \frac{\eta(\mathbf{e})}{\eta^\ell(\mathbf{e})},
\end{align}

\noindent and can be derived geometrically (see \figurename~\ref{fig:sawtooth}).
Note that we use the $\max$ in~\eqref{eq:sawtooth_def} so as to obtain the lowest (i.e., best) upper bound. We now rewrite the sawtooth projection in a simpler form:~
\begin{align}\label{eq:sawtooth_redef}
    \textbf{swt}(\Upsilon_k,\!\eta)\! &\stackrel{(a)}{=} \! y^0(\eta) \!+\! \min_{\ell \in \mathcal{L}} \!\Big\{(v^\ell \!-\!y^0(\eta^\ell)) \min_{\mathbf{e}:\eta^\ell(\mathbf{e})>0} \frac{\eta(\mathbf{e})}{\eta^\ell(\mathbf{e})}\Big\} \nonumber\\
     & \stackrel{(b)}{=}\! y^0(\eta) \!+\! \min_{\ell \in \mathcal{L}} \max_{\mathbf{e}:\eta^\ell(\mathbf{e})>0} \!\Big\{\frac{\eta(\mathbf{e})}{\eta^\ell(\mathbf{e})} (v^\ell \!-\! y^0(\eta^\ell)) \Big\} \nonumber \\
     & \stackrel{(c)}{=}\! \min_{\ell \in \mathcal{L}} \Big\{y^0(\eta) \!+\! \max_{\mathbf{e}:\eta^\ell(\mathbf{e})>0} \!\Big\{\frac{\eta(\mathbf{e})}{\eta^\ell(\mathbf{e})} (v^\ell \!-\! y^0(\eta^\ell)) \Big\}\Big\} \nonumber \\
     & \stackrel{(d)}{=}\! \min_{\ell \in \mathcal{L}} \textbf{swt}^\ell(\Upsilon_k,\eta).
\end{align}

\noindent Step $(a)$ coincides with Definition~\eqref{eq:sawtooth_def}; step $(b)$ holds because $v^\ell$ is not greater than $y^0(\eta^\ell)$, since $y^0(\eta^\ell)$ represents the interpolation using only the corner points (see \figurename~\ref{fig:sawtooth}); in step $(c)$, we move $y^0(\eta)$ inside the $\min_\ell$, since it does not depend on $\ell$; finally, we define $\textbf{swt}^\ell(\Upsilon_k,\eta)$ in step $(d)$.

\noindent We also introduced $\textbf{swt}^\ell(\Upsilon_k,\eta)$, which will be used in the next subsection, as the sawtooth projection obtained using the $\ell$-th element of $\Upsilon_k$.

The sawtooth projection produces higher (i.e., worse) upper bounds than the convex hull projection and thus  MPS may require more iterations to converge (however, a single iteration can be performed much more quickly), but convergence is still guaranteed~\cite{Dibangoye2014}.

\subsection{Constraint Programming Formulation}\label{subsec:CP}

The key step to perform to find the policy is solving the $\max$ in~\eqref{eq:v_beta_C}. Although this would be possible by performing the \emph{exhaustive backup}, i.e., by inspecting all possible choices of $\boldsymbol{\sigma}$, it is more practical to introduce faster solutions.
Constraint Programming~\cite{Dechter2003} is a technique to express hard and soft constraints as an optimization problem. We now use it to reformulate~\eqref{eq:v_beta_C} with $C = \textbf{swt}$.

Using the notation of Dibangoye \emph{et al.}~\cite{Dibangoye2014}, we define~
\begin{align}\label{eq:W}
    W_k(\eta_k,\boldsymbol{\sigma},\ell) \triangleq \phi_k(\eta_k,\boldsymbol{\sigma}) + \textbf{swt}^\ell(\Upsilon_k,\omega(\eta_k,\boldsymbol{\sigma})).
\end{align}

Combining~\eqref{eq:sawtooth_redef} and~\eqref{eq:W}, we can rewrite Equation~\eqref{eq:v_beta_C} as~
\begin{align}
    \bar{v}_k(\eta_k) &= \max_{\boldsymbol{\sigma}} \{\phi_k(\eta_k,\boldsymbol{\sigma}) + \textbf{swt}(\Upsilon_k,\!,\omega(\eta_k,\boldsymbol{\sigma}))\} \nonumber \\
    &= \max_{\boldsymbol{\sigma}} \{ \phi_k(\eta_k,\boldsymbol{\sigma}) + \min_{\ell \in \mathcal{L}} \textbf{swt}^\ell(\Upsilon_k,\omega(\eta_k,\boldsymbol{\sigma})) \} \nonumber \\
    &=\max_{\boldsymbol{\sigma}} \min_{\ell \in \mathcal{L}} W_k(\eta_k,\boldsymbol{\sigma},\ell).
\end{align}

To solve the previous equation, we split it in $|\mathcal{L}|$ separate mixed-integer programs:~
\begin{align}\label{eq:bar_v_L}
    &\bar{v}_k(\eta_k) = \max_{\ell \in \mathcal{L}} \bar{v}_k^\ell(\eta_k),
\end{align}

\noindent with~
\begin{subequations}
\label{eq:bar_v_L_ell}
\begin{align}
    &\bar{v}_k^\ell(\eta_k) \triangleq \max_{\boldsymbol{\sigma}}  W_k(\eta_k,\boldsymbol{\sigma},\ell), \label{eq:v_beta_max}\\
    &\text{subject to:} \quad W(\eta_k,\boldsymbol{\sigma},\ell) \leq W_k(\eta_k,\boldsymbol{\sigma},l), \quad \forall l \in \mathcal{L}. \label{eq:st_W_W}
\end{align}
\end{subequations}

\emph{\textbf{Weighted Constraint Satisfaction Problem (WCSP).}} Focus now on the optimization $\max_{\boldsymbol{\sigma}}  W(\eta_k,\boldsymbol{\sigma},\ell)$ \emph{without} any constraints. This can be formulated as a WCSP as follows. First, we rewrite $W(\eta_k,\boldsymbol{\sigma},\ell)$ using~\eqref{eq:rho_def}, \eqref{eq:phi_def}, \eqref{eq:y_0_def}, \eqref{eq:sawtooth_redef}, and~\eqref{eq:W}:~
\begin{align}
    & W_k(\eta_k,\boldsymbol{\sigma},\ell) = (1-\beta_0)^k \sum_{\mathbf{e} \in \boldsymbol{\mathcal{E}}} \eta_k(\mathbf{e}) r(\boldsymbol{\sigma}(\mathbf{e}))  \\
    & + (1-\beta_0)^k \sum_{\mathbf{e}^\prime \in \boldsymbol{\mathcal{E}}} \eta_{k+1}(\mathbf{e}^\prime) z^{{\rm iter}-1}(\mathbf{e}^\prime) \nonumber \\
    & + \!\sum_{\mathbf{e}^\prime \in \boldsymbol{\mathcal{E}}} \! \eta_{k+1}(\mathbf{e}^\prime) \Upsilon_k(\mathbf{e}^\prime) \! + \!\!\max_{\mathbf{e}^{\prime\prime}:\eta^\ell(\mathbf{e}^{\prime\prime})>0} \! \Big\{\! \frac{\eta_{k+1}(\mathbf{e}^{\prime\prime})}{\eta^\ell(\mathbf{e}^{\prime\prime})} (v^\ell \! - \! y^0(\eta^\ell)) \Big\}, \nonumber
\end{align}

\noindent where $\eta_{k+1} = \omega(\eta_k,\boldsymbol{\sigma})$, and we used $\mathbf{e}$, $\mathbf{e}^\prime$ and $\mathbf{e}^{\prime\prime}$ to differentiate the indices. Note that, since $\ell$ is given, the term $y^0(\eta^\ell)$ is fixed. Using the occupancy update formula of Equation~\eqref{eq:eta_k_update}, we get~
\begin{align}
    &W(\eta_k,\boldsymbol{\sigma},\ell) = (1-\beta_0)^k  \sum_{\mathbf{e} \in \boldsymbol{\mathcal{E}}} \eta_k(\mathbf{e}) r(\boldsymbol{\sigma}(\mathbf{e})) \label{eq:W_WCSP}\\
    & + (1-\beta_0)^k \sum_{\mathbf{e}^\prime \in \boldsymbol{\mathcal{E}}} \sum_{\mathbf{e} \in \boldsymbol{\mathcal{E}}} \eta_k(\mathbf{e}) p(\mathbf{e}^\prime|\mathbf{e},\boldsymbol{\sigma}(\mathbf{e})) z^{{\rm iter}-1}(\mathbf{e}^\prime) \nonumber\\
    & + \sum_{\mathbf{e}^\prime \in \boldsymbol{\mathcal{E}}} \sum_{\mathbf{e} \in \boldsymbol{\mathcal{E}}} \eta_k(\mathbf{e}) p(\mathbf{e}^\prime|\mathbf{e},\boldsymbol{\sigma}(\mathbf{e})) \Upsilon_k(\mathbf{e}^\prime) \nonumber\\
    & + \max_{\mathbf{e}^{\prime\prime}:\eta^\ell(\mathbf{e}^{\prime\prime})>0} \Big\{ \sum_{\mathbf{e} \in \boldsymbol{\mathcal{E}}} \frac{\eta_k(\mathbf{e}) p(\mathbf{e}^{\prime\prime}|\mathbf{e},\boldsymbol{\sigma}(\mathbf{e})) }{\eta^\ell(\mathbf{e}^{\prime\prime})} (v^\ell-y^0(\eta^\ell)) \Big\}. \nonumber
\end{align}

\noindent Since the first terms of~\eqref{eq:W_WCSP} do not depend on $\mathbf{e}^{\prime\prime}$, we move them inside the $\max_{\mathbf{e}^{\prime\prime}}$ and take the common sum over $\mathbf{e}$; then, we note that all terms multiply $\eta_k(\mathbf{e})$. Thus, by introducing a variable $w_\ell(\cdot)$, we obtain~
\begin{align}
    &W(\eta_k,\boldsymbol{\sigma},\ell) = \max_{\mathbf{e}^{\prime\prime}:\eta^\ell(\mathbf{e}^{\prime\prime})>0}\sum_{\mathbf{e} \in \boldsymbol{\mathcal{E}}} \eta_k(\mathbf{e}) w_\ell(\mathbf{e},\boldsymbol{\sigma}(\mathbf{e}),\mathbf{e}^{\prime\prime}).
\end{align}

For every fixed $\ell$, the WCSP is formally defined as follows. The variables are defined by $\boldsymbol{\sigma}$ as in~\eqref{eq:sigma_struct} (i.e., the actions $\mathbf{a} \in \boldsymbol{\mathcal{A}}$ for every $\mathbf{e} \in \boldsymbol{\mathcal{E}}$) plus the index $\mathbf{e}^{\prime\prime}$. The domains are the same as in the original problem, i.e., $\boldsymbol{\mathcal{A}}$ for $\boldsymbol{\sigma}$ and $\boldsymbol{\mathcal{E}}$ for $\mathbf{e}^{\prime\prime}$. A WCSP is fully specified by its constraints, which are of the form~
\begin{align}
    \textbf{constraint}_\ell(\mathbf{e}) = M - \eta_k(\mathbf{e}) w_\ell(\mathbf{e},\boldsymbol{\sigma}(\mathbf{e}),\mathbf{e}^{\prime\prime}).
\end{align}

\noindent The total number of constraints is at most $|\boldsymbol{\mathcal{E}}|$, one for every possible battery level. $M$ is a large number used to cast the WCSP in its standard form. Standard WCSP solvers compute the following quantity:~
\begin{align}
    \min_{\sigma_k^i(e^i), \, \forall e^i, \, \forall i, \ {\rm and}\ \mathbf{e}^{\prime\prime} \in \boldsymbol{\mathcal{E}} }
        \ \sum_{\mathbf{e} \in \boldsymbol{\mathcal{E}}} 
        \textbf{constraint}_\ell(\mathbf{e}),
\end{align}

\noindent whose solution is equal to the solution of~\eqref{eq:v_beta_max}.
In practice, for every $\mathbf{e} \in \boldsymbol{\mathcal{E}}$, the quantity $\textbf{constraint}_\ell(\mathbf{e})$ is evaluated for all the combinations of $\boldsymbol{\sigma}(\mathbf{e}) \in \boldsymbol{\mathcal{A}}$ and $\mathbf{e}^{\prime\prime} \in \boldsymbol{\mathcal{E}}$ independently of the others constraints. Then, they are summed together and the minimum among all the solutions is chosen. However, a solution referred to a constraint $\textbf{constraint}_\ell(\mathbf{e})$ may be related to other solutions. For example, different states $\mathbf{e}$ have some common entries (e.g., both $\langle 0,0 \rangle \in \boldsymbol{\mathcal{E}}$ and $\langle 0,1 \rangle \in \boldsymbol{\mathcal{E}}$ have the first entry equal to $0$); in this case, the corresponding actions must have some common element even if they are referring to different constraints (e.g., $\boldsymbol{\sigma}(\langle 0,0 \rangle) = \langle \sigma^0(0), \sigma^1(0) \rangle$ and $\boldsymbol{\sigma}(\langle 0,1 \rangle) = \langle \sigma^0(0), \sigma^1(1) \rangle$ must have the first entry in common).

The main advantage of using a WCSP solver is that the decentralized policy does not need to be computed as a whole, but can be divided in constraints which are later combined together. 

\medskip 

So far, we have only focused on~\eqref{eq:v_beta_max} (i.e., on a single $\ell$). To solve~\eqref{eq:bar_v_L}, we need to compute $|\mathcal{L}|$ different WCSPs (one for every $\ell \in \mathcal{L}$) and take the maximum among all solutions.
However this approach presents one major drawback: for a fixed $\ell$, solving a single WCSP may not be sufficient. Indeed, the solution must also satisfy~\eqref{eq:st_W_W}, which has been completely neglected in the definition of the WCSP. If a constraint were violated, then the solution of the WCSP would be infeasible and should be discarded. In this case, Dibangoye~\emph{et al.} (see~\cite[Section~3.4.2]{Dibangoye2013}) propose to formulate a new WCSP in which the previous infeasible solution results in a very high cost constraint (and thus it is never chosen in the solution process). If the new solution also turns out to be infeasible, the procedure is repeated. The iterations stop if a feasible solution is found or if all decentralized actions have been examined.

Although the previous method formally leads to the optimal solution, it may often degenerate in an exhaustive search (i.e, examining all the decentralized policies). The corresponding complexity would be $\mathcal{O}((S_a)^{e_{\rm max}\times N})$ if all users had the same battery size $e_{\rm max}$ (see the structure of $\boldsymbol{\sigma}(\mathbf{e})$ in Equation~\eqref{eq:sigma_struct}), i.e., exponential in $N$. This operation is computationally infeasible when lots of possibilities are involved.
Thus, optimally solving a Dec-MDP with guarantees on the worst case performance is still an open issue. In the next section, we propose two suboptimal approaches for handling the problem based on the previous results.

\section{Sub-Optimal Solutions of the Internal Layer}\label{sec:sub_opt_solutions}

Since the main issue of the exhaustive search is that the space of variables in Problem~\eqref{eq:bar_v_L}-\eqref{eq:bar_v_L_ell} is exceedingly large, we aim at reducing this space, so that $\boldsymbol{\sigma}$ cannot take all possible values but is constrained to lie in a smaller subset. The problem now is to define such a subset. In the next subsection we present an approach based on WCSPs, whereas in Section~\ref{subsec:sub_opt_parametric} we introduce a different scheme based on parametric policies.

\subsection{WCSP-Based Policies}\label{subsec:sub_opt_WCSP}

In this case, we exploit the results about WCSPs presented in Section~\ref{subsec:CP} to find a suboptimal policy. The proposed algorithm is as follows.
\begin{algorithm}[H]
\caption{(Suboptimal policy using WCSPs)}\label{alg:WCSP}
\begin{algorithmic}[1]
\For{$\ell \in \mathcal{L}$} \label{alg:WCSP:every_ell}
    \State $\boldsymbol{\sigma}_\ell$ $\gets$ Solve WCSP for $\ell$ given $\eta_k$ \label{alg:WCSP:sigma_ell}
    \For{$l \in \mathcal{L}$} \label{alg:WCSP:l}
        \State Evaluate $W(\eta_k,\boldsymbol{\sigma}_\ell,l)$ \label{alg:WCSP:W}
    \EndFor
    \State $l^\star \gets \arg\min_{l \in \mathcal{L}} W(\eta_k,\boldsymbol{\sigma}_\ell,l)$ \label{alg:WCSP:min}
    \State $\bar{v}_k^\ell(\eta_k) \gets W(\eta_k,\boldsymbol{\sigma}_\ell,l^\star)$ \label{alg:WCSP:v_beta}
\EndFor
\State $\ell^\star \gets \max_\ell \bar{v}_k^\ell(\eta_k)$ \label{alg:WCSP:max} \\
\Return $\boldsymbol{\sigma}_{\ell^\star}$ \label{alg:WCSP:return}
\end{algorithmic}
\end{algorithm}

As required by the $\max$ in~\eqref{eq:bar_v_L}, we look at every $\ell \in \mathcal{L}$ (Line~\ref{alg:WCSP:every_ell}) and, at the end of the algorithm, we return the solution with the maximum value (Lines~\ref{alg:WCSP:max}-\ref{alg:WCSP:return}). Lines~\ref{alg:WCSP:sigma_ell}-\ref{alg:WCSP:v_beta} solve Problem~\eqref{eq:bar_v_L_ell} suboptimally as described in the following.

First, we solve the WCSP for a fixed $\ell$ (i.e., we solve~\eqref{eq:v_beta_max}), and find the corresponding solution $\boldsymbol{\sigma}_\ell$. Then, using $\boldsymbol{\sigma}_\ell$, we evaluate $W(\eta_k,\boldsymbol{\sigma}_\ell,l)$ for every index $l$. Two cases should now be examined, which can be handled in a fully equivalent way, but have different meanings:
\begin{itemize}
    \item If $W(\eta_k,\boldsymbol{\sigma}_\ell,\ell) \leq W(\eta_k,\boldsymbol{\sigma}_\ell,l),\, \forall l \in \mathcal{L}$, then $\boldsymbol{\sigma}_\ell$ would be an optimal solution of~\eqref{eq:bar_v_L_ell}, since it maximizes~\eqref{eq:v_beta_max} (Line~\ref{alg:WCSP:sigma_ell}) and satisfies~\eqref{eq:st_W_W}. In this case, $\ell \equiv l^\star$ (Line~\ref{alg:WCSP:min}), and $\bar{v}_k^\ell(\eta_k)$ is the solution of~\eqref{eq:bar_v_L_ell};
    \item Instead, when there exists $l \neq \ell$ such that $W(\eta_k,\boldsymbol{\sigma}_\ell,l) < W(\eta_k,\boldsymbol{\sigma}_\ell,\ell)$, then $\boldsymbol{\sigma}_\ell$ \emph{is not optimal} for index $\ell$. In this case, the optimal approach would require the execution of a new WCSP discarding the previous solution (in the new WCSP, $\boldsymbol{\sigma}_\ell$ would become a very high cost solution). Instead, in Algorithm~\ref{alg:WCSP}, we implicitly make the following observation: $\boldsymbol{\sigma}_\ell$ is a feasible solution (i.e., it satisfies~\eqref{eq:st_W_W}) of $\bar{v}_k^{l^\star}(\eta_k)$, where $l^\star$ is such that $W(\eta_k,\boldsymbol{\sigma}_\ell,l^\star) \leq W(\eta_k,\boldsymbol{\sigma}_\ell,l),\, \forall l \in \mathcal{L}$ (Line~\eqref{alg:WCSP:min}). Therefore, solution $W(\eta_k,\boldsymbol{\sigma}_\ell,l^\star)$ is \emph{feasible}; for simplicity, we improperly save its value in $\bar{v}_k^\ell(\eta_k)$ (Line~\ref{alg:WCSP:v_beta}). By doing so, at the end of Line~\ref{alg:WCSP:v_beta}, $\bar{v}_k^\ell(\eta_k)$ does not represent the solution of~\eqref{eq:bar_v_L_ell} associated to index $\ell$, but it contains a feasible solution for some index $l^\star$.
\end{itemize}

In practice, while executing Algorithm~\ref{alg:WCSP}, the space of variables of Problem~\eqref{eq:bar_v_L} is defined by the solutions of WCSPs at every iteration (thus, it is not determined a priori).

The proposed approach is faster than the optimal one, since it completely avoids the exhaustive search; however, in general, it is suboptimal and thus achieves worse performance.

\subsection{Parametric Policies}\label{subsec:sub_opt_parametric}

Another possibility to avoid the exhaustive search step is to use parametric policies and thus reduce the number of optimization variables to few parameters.
In particular, we force the actions of user $i$ to follow a predetermined structure:~
\begin{align}
    \sigma^i(e^i) = f_{\rm par}^i(\Theta^i,e^i)
\end{align}

\noindent where $e^i$ is the independent variable and $\Theta^i$ is a set of parameters which specify the structure of $f_{\rm par}^i$. For example, if we used $\Theta^i = \{\theta^i\}$, and a simple linear function $f_{\rm par}^i(\Theta^i,e^i) = \theta^i e^i$, the only optimization variable of user $i$ would be $\theta^i$, and not $\sigma^i(0),\ldots,\sigma^i(e_{\rm max}^i)$ as in the original problem.
In this case, for a symmetric scenario, the complexity of the exhaustive search step goes from $\mathcal{O}((S_a)^{e_{\rm max} \times N\vphantom{|}})$ to $\mathcal{O}((S_\theta)^{N\vphantom{!}})$, therefore it remains exponential in $N$ but with a much smaller coefficient in the exponent. $S_\theta$ is the number of values that $\theta^i$ can assume.\footnote{We note that, although we did not reduce the theoretical complexity of the exhaustive search (which is still exponential), using smaller coefficients allows much faster numerical evaluations. The more general problem of developing heuristic schemes with lower complexity is still open.} 

In our scenario we force $f_{\rm par}^i(\Theta^i,e^i)$ to be a non-decreasing function of $e^i$ as in~\cite{Michelusi2013}, which implies that higher energy levels cannot correspond to lower transmission probabilities.

\section{Numerical Evaluation}\label{sec:num_eval}

\begin{figure*}[t]
    \centering
    \begin{minipage}[b]{.49\textwidth}
        \centering
        \includegraphics[trim = 0mm 0mm 0mm 0mm,  clip, width=0.98\columnwidth]{txProb_s01_Pe1.eps}
    \end{minipage}%
    \hfill%
    \begin{minipage}[b]{.49\textwidth}
        \centering
        \includegraphics[trim = 0mm 0mm 0mm 0mm,  clip, width=0.99\columnwidth]{txProb_s81_Pe1.eps}
    \end{minipage}\\
    \vspace{.2cm}
    \centering
    \begin{minipage}[b]{.49\textwidth}
        \centering
        \includegraphics[trim = 0mm 0mm 0mm 0mm,  clip, width=1\columnwidth]{txProb_s73_Pe1.eps}
    \end{minipage}%
    \hfill%
    \begin{minipage}[b]{.49\textwidth}
        \centering
        \includegraphics[trim = 0mm 0mm 0mm 0mm,  clip, width=1\columnwidth]{txProb_s09_Pe1.eps}
    \end{minipage}\\
    \hspace{.005\textwidth}
    \begin{minipage}[t]{0.999\textwidth}
        \caption{Transmission probabilities as a function of time for two users when $\beta_0 = 0.05$ and $p_B^1 = p_B^2 = 0.1$ for different initial battery levels.}
        \label{fig:txProb_Pe1}
    \end{minipage}
    \vspace{-.5cm}
\end{figure*}

\begin{figure*}[t]
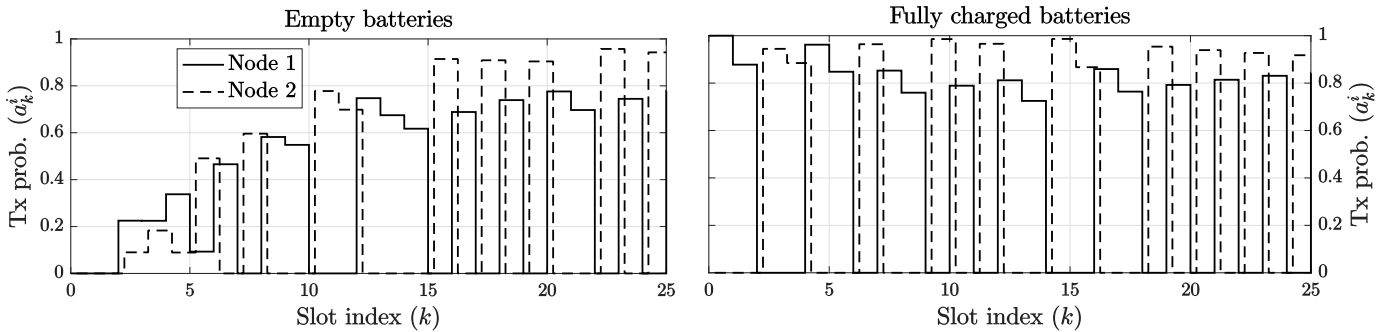

    \centering
    \begin{minipage}[b]{.49\textwidth}
        \centering
        \includegraphics[trim = 0mm 0mm 0mm 0mm,  clip, width=1\columnwidth]{txProb_s01_Pe9.eps}
    \end{minipage}%
    \hfill%
    \begin{minipage}[b]{.49\textwidth}
        \centering
        \includegraphics[trim = 0mm 0mm 0mm 0mm,  clip, width=1\columnwidth]{txProb_s81_Pe9.eps}
    \end{minipage}\\
    \hspace{.005\textwidth}
    \begin{minipage}[t]{0.999\textwidth}
        \caption{Transmission probabilities as a function of time for two users when $\beta_0 = 0.05$ and $p_B^1 = p_B^2 = 0.9$ for different initial battery levels.}
        \label{fig:txProb_Pe8}
    \end{minipage}
    \vspace{-.5cm}
\end{figure*}

The numerical evaluation is performed using two nodes, since the complexity grows super-exponentially with the number of users. Indeed the size of the occupancy state evolves exponentially with $N$, and the exhaustive search operation (exponential in $N$), or a suboptimal approach, is to be performed for every element of the occupancy state. If not otherwise stated, we adopt the following parameters: the batteries can contain up to $e_{\rm max}^1 = e_{\rm max}^2 = 8$ energy quanta; the energy arrival processes are i.i.d. over time and the probability of receiving one energy quantum is $p_B^1 = p_B^2$ in every slot; when a transmission is performed a reward $V^i = \ln(1+\Lambda^i H^i)$ is accrued, where $V^i$ represents the normalized transmission rate in a slot, and $H^i$ is an exponentially distributed random variable with mean $1$ (see~\cite{Michelusi2012}); the transmission probabilities in $[0,1]$ are uniformly quantized with $S_a = 19$ samples; the average normalized SNRs are $\Lambda^1 = 6$ and $\Lambda^2 = 3$; both devices have the same weight; to perform a transmission $m^1 = m^2 = 2$ energy quanta are drawn from the battery; finally, $\beta$ has period $q = 1$ (i.e., it is constant over time). All the numerical evaluations were written in C++ and, for solution of the weighted constraint satisfaction problems, we used ToulBar2~\cite{toulbar2}, a highly efficient solver of WCSPs. We first focus on the solution of the internal layer (i.e., we only look at $R_0(\pi_{\mathbf{e}_0},\mathbf{e}_0)$), and discuss later how the external layer performs.

\emph{\textbf{Transmission Probabilities.}}
In \figurename s~\ref{fig:txProb_Pe1} (low energy arrival rates) and~\ref{fig:txProb_Pe8} (high energy arrival rates) we show the transmission probabilities of the parametric decentralized policy of Section~\ref{subsec:sub_opt_parametric}, where $f_{\rm par}$ is a linear function, $\Theta^i = \{\theta^i\}$ and $\theta^i$ is such that $\theta^i e_{\rm max}^i \in \mathcal{A}^i$. The dashed lines have been slightly manually shifted to the right only for graphical purposes. In these figures, we only focus on the internal layer, thus we study the system behavior between two consecutive SYNC slots.

The main difference between \figurename s~\ref{fig:txProb_Pe1} and~\ref{fig:txProb_Pe8} is that, after many slots, the transmission probabilities are both greater than zero in the first case, whereas an almost pure time-orthogonal approach is used in the latter, regardless of the initial energy levels. 
Thus, when the energy resources are scarce (i.e., low energy arrivals) then, if $k$ is large, an orthogonal scheme in which collisions are avoided is suboptimal. The trade-off between orthogonal and random access schemes can be intuitively explained as follows. As time goes on, nodes lose information about the global state of the system, thus a device does not know the energy level of the other. In this case, an orthogonal scheme might be highly inefficient: if a slot were assigned only to Node~$1$, but this did not have enough energy for transmission, then the slot would be unused. Since the energy resources are scarce, it is likely that such a case happens. Instead, when both have a lot of energy, it is easier to estimate the energy level of the nodes, and thus pre-assigning the slots is possible (and in fact optimal). This means that, in the low energy case, it is better to transmit even if the collision probability is non-negligible, to compensate the (likely) lack of energy of the other user.
This represents the main novelty of EH scenarios over traditional ones in which, as explained in Section~\ref{subsec:reward}, an orthogonal decentralized access scheme is optimal.

The same effect can be observed in \figurename~\ref{fig:txProb_Pe1} when the initial energy levels are high. Indeed, when $k$ is small, nodes still have information about the state of the other device; therefore, an orthogonal approach is optimal in this case also, and collisions are avoided since it is very likely that a node has enough energy to perform the transmission.

Note that when an orthogonal access scheme is employed, Node~$1$ is advantaged with respect to Node~$2$ (this can be clearly seen in \figurename~\ref{fig:txProb_Pe8}, where more slots are allocated to the first node). Indeed, the SNR of the first node ($\Lambda^1$) is greater than the SNR of the other ($\Lambda^2$), and therefore higher returns are obtained by Node~$1$ when a transmission is performed. In this case, fairness is not achieved because of the near-far effect (a node with a better channel is advantaged over the other); however, the network could be rebalanced by changing the weights $w^1$, $w^2$.

Finally, note that in \figurename~\ref{fig:txProb_Pe1} the average transmission probabilities in the long run almost coincide with the energy arrival rate divided by $m^i$, so as to achieve energy neutrality.

\begin{figure}[!t]
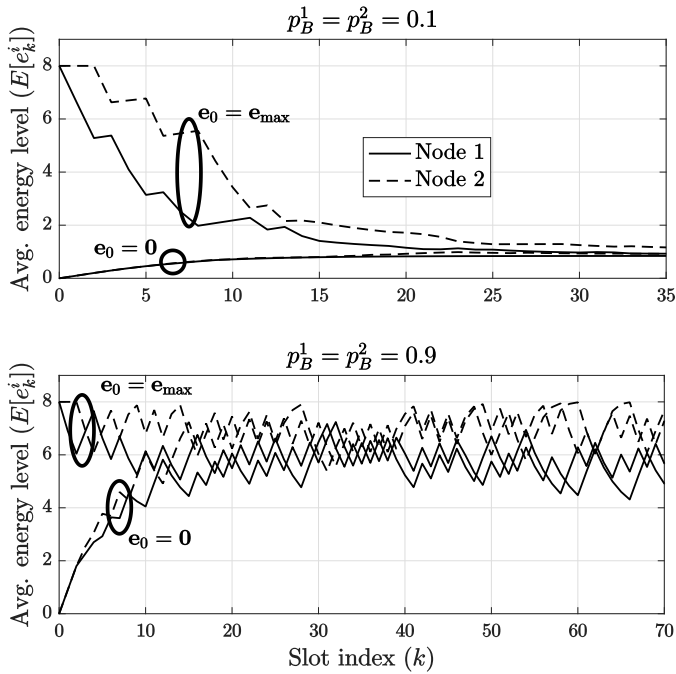

  \centering
  \includegraphics[trim = 0mm 0mm 0mm 0mm, width=1\columnwidth]{energyLevels_Pe1.eps}\\
    \vspace{.5cm}
  \includegraphics[trim = 0mm 0mm 0mm 0mm, width=1\columnwidth]{energyLevels_Pe9.eps}
  \caption{Battery level evolution as a function of time for two users when $\beta_0 = 0.05$, and $p_B^1 = p_B^2 = 0.1$ or $0.9$ for different initial battery levels.}
  \label{fig:energyLevels}
\end{figure}

\emph{\textbf{Energy Levels.}} \figurename~\ref{fig:energyLevels} shows another interesting, though predictable, result: despite the initial energy level, in the long run all the energy levels of the same device converge, approximately, to the same value. This is because all the initial fluctuations have been absorbed by the batteries. Note that the energy levels of Node~2 are higher because $\Lambda^2 < \Lambda^1$, thus Node~2 transmits less frequently than the other and consumes less energy, on average.

\begin{figure}[t]
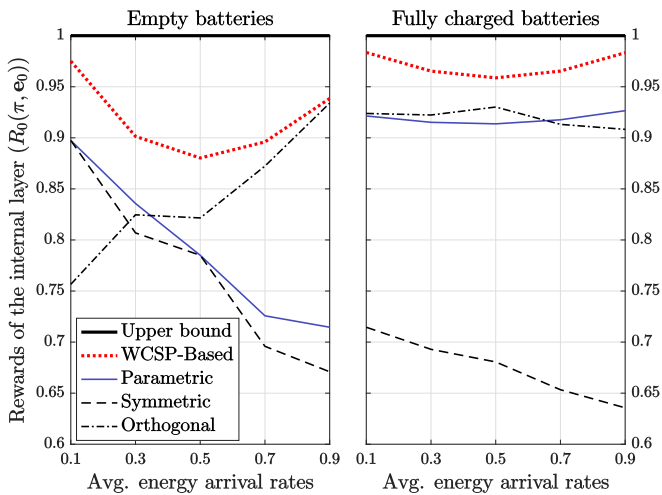

    \centering
    \begin{minipage}[b]{.5\columnwidth}
        \centering
        \includegraphics[trim = 0mm 0mm 0mm 0mm,  clip, width=1\columnwidth]{rewards_internals_1.eps}
    \end{minipage}%
    \hfill%
    \begin{minipage}[b]{.5\columnwidth}
        \centering
        \includegraphics[trim = 0mm 0mm 0mm 0mm,  clip, width=.910\columnwidth]{rewards_internals_81.eps}
    \end{minipage}\\
    \hspace{.005\columnwidth}
    \begin{minipage}[t]{0.999\columnwidth}
        \caption{Long-term rewards of the internal layer as a function of the energy arrival rates $p_B^1=p_B^2$ for two users when $\beta_0 = 0.2$ for batteries initially empty (left) or fully charged (right).}
        \label{fig:rewards_internal}
    \end{minipage}
    \vspace{-.5cm}
\end{figure}

\emph{\textbf{Internal Reward.}} In \figurename~\ref{fig:rewards_internal} we show the long-term discounted reward as a function of the energy arrival rate for the decentralized scheme solved using WCSPs (Section~\ref{subsec:sub_opt_WCSP}), the decentralized parametric scheme (Section~\ref{subsec:sub_opt_parametric}), a fully orthogonal approach, and a fully symmetric scheme. The curves are normalized with respect to the upper bound, given by the centralized scheme of Equation~\eqref{eq:R_beta_cent} (which, however cannot be achieved in decentralized scenarios, in general). To understand the trend of the curves, it is important to remark that the first slots after the initial SYNC slot are the most important ones for two reasons:
\begin{enumerate}
    \item The decentralized reward $\phi_k(\cdot)$ defined in~\eqref{eq:phi_def} decreases with $k$. Therefore, the initial slots have higher weights and contribute more to the global reward;
    \item There is more information about the state of the other device in the initial slots.
\end{enumerate}

Therefore, when the initial batteries are fully charged, the centralized and decentralized schemes are closer. Indeed, since $\mathbf{e}_0 = \mathbf{e}_{\rm max}$, decentralized and centralized rewards in the first slots are almost equivalent.
Instead, if the batteries are initially fully discharged, the gap between centralized and decentralized is wider. In this case, the first slots do not play a fundamental role, since there is not a lot of energy to exploit. Therefore, most of the gain is obtained for higher $k$, which in turn leads to a less informative situation about the state of the other device. In this case, the centralized scheme may perform much better than the decentralized one because of this lack of information.

Although it is not visible from \figurename~\ref{fig:rewards_internal} because of the normalization, the curves increase with $p_B^1=p_B^2$, because more energy can be harvested (see~\cite[Section~V]{Biason2017WCNC}). The lowest normalized reward for the WCSP-based policy is obtained around $p_B^1 = p_B^2 = 0.5$; indeed, this corresponds to the least informative case, since the battery fluctuations are not predictable at all. 


Finally, note that the decentralized policy obtained using WCSP outperforms the parametric policy in almost every case. However, this is also strongly influenced by the number of parameters $\Theta^i$ we used, and using more parameters would lead to better performance. Moreover, it can be clearly seen that the fully orthogonal and the symmetric policies are strongly suboptimal in this scenario, therefore using an optimized approach significantly improves the throughput of the network.
An interesting outcome of \figurename~\ref{fig:rewards_internal} is that the gap between centralized and decentralized schemes may be small ($< 15\%$), when the decentralized policy is correctly designed. Therefore, using low values of $\beta$ (e.g., $\beta = 0.05$) we are able to achieve the twofold goal of greatly reducing the overhead in the network, since the policy is distributed only during the SYNC slots, and of achieving good performance.

\begin{figure}[!t]
  \centering
  \includegraphics[trim = 0mm 0mm 0mm 0mm, width=1\columnwidth]{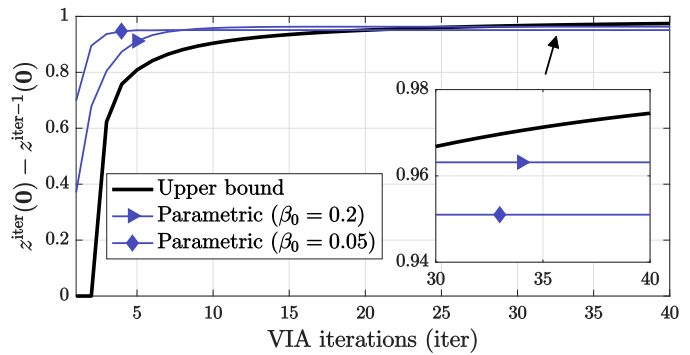}
  \caption{Transmission probabilities as a function of time for two users with batteries initially empty when $p_B^1 = b_B^2 = 0.05$.}
  \label{fig:VIA_iter}
\end{figure}

\begin{figure}[!t]
  \centering
  \includegraphics[trim = 0mm 0mm 0mm 0mm, width=1\columnwidth]{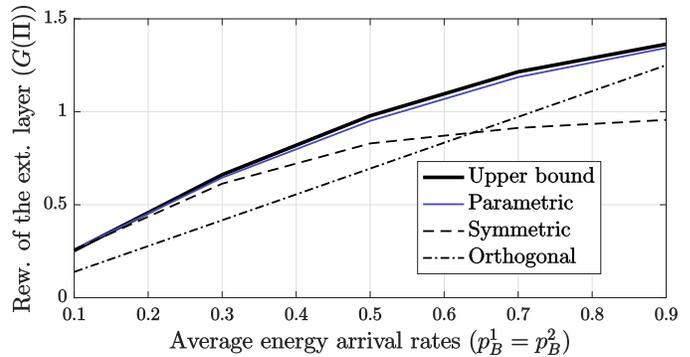}
  \caption{Long-term rewards as a function of the energy arrival rates for two users when $p_B^1 = b_B^2 = 0.05$.}
  \label{fig:rewards_external}
\end{figure}

\emph{\textbf{External Reward.}} 
We now describe the performance of the complete system, i.e., of the external layer. First, we show the iterations of VIA (Equation~\eqref{eq:Bellman_external}) in \figurename~\ref{fig:VIA_iter}. Note that the $y$-axis represents the difference between two consecutive steps of VIA, which converges to $G(\Pi^\star)$ (see the Relative Value Iteration Algorithm in~\cite[Vol.~1, Sec.~7.4]{Bertsekas2005}).
The case iter~$=1$ corresponds to the internal layer only (i.e., $\beta_0 \cdot R_0(\pi_{\mathbf{0}},\mathbf{0})$), since $z^0(\mathbf{e}) = 0,\ \forall \mathbf{e}\in \boldsymbol{\mathcal{E}}$. We highlight that, to solve a single step of VIA, many decentralized optimization steps are performed (one for every $\mathbf{e} \in \boldsymbol{\mathcal{E}}$). From the figure, it can be clearly seen that only few iterations are required for convergence, especially for lower $\beta_0$ (which are the most computationally expensive, since the intervals between SYNC slots are longer). Note that, when VIA converges, the higher $\beta_0$, the higher the reward, as expected; moreover, the upper bound curve (i.e., the centralized case) outperforms the others (in this case, there is less uncertainty and more global information).
In \figurename~\ref{fig:rewards_external}, we plot the reward $G(\Pi)$ of the external layer as a function of the energy arrival rates. The values reported here are derived from the last step of VIA. As expected, $G(\Pi)$ increases with $p_B^1 = p_B^2$ (more energy available). It is interesting to observe that the parametric policy is very close to the upper bound, whereas the rewards of the other, simpler policies are much lower.

\section{Conclusions}\label{sec:conclusions}

In this paper, we studied a decentralized optimization framework for an energy harvesting communication network with collisions. We used a multi-layer Markov setup to model the system. An external layer, that models the jumps between SYNC slots (the time instants at which the policy is computed) is optimized as a Markov Decision Process (MDP) whose actions are given by an internal layer, modeled as a decentralized-MDP. To solve the external layer we used the Value Iteration Algorithm, whereas the Markov Policy Search algorithm was employed for the internal layer. Because of the exponential complexity of the exhaustive search used in the optimal policy, we presented two simpler schemes, namely WCSP-based and parametric approaches. In our numerical evaluations we described the trade-off between accessing the channel and energy arrivals, and we showed that a pure orthogonal access mechanism is suboptimal under harvesting constraints. Moreover, we noted that decentralized schemes, if correctly designed, may achieve high performance while significantly reducing the signaling overhead in the network.

Our work introduces a principle of optimality in decentralized energy harvesting networks. However, because of the complexity burden of the optimal policy, additional studies are required for proposing  more practical schemes which inherit the key properties of our framework while being less computationally demanding.

\bibliography{../../../bibliography}{}

\begin{thebibliography}{10}
\providecommand{\url}[1]{#1}
\csname url@samestyle\endcsname
\providecommand{\newblock}{\relax}
\providecommand{\bibinfo}[2]{#2}
\providecommand{\BIBentrySTDinterwordspacing}{\spaceskip=0pt\relax}
\providecommand{\BIBentryALTinterwordstretchfactor}{4}
\providecommand{\BIBentryALTinterwordspacing}{\spaceskip=\fontdimen2\font plus
\BIBentryALTinterwordstretchfactor\fontdimen3\font minus
  \fontdimen4\font\relax}
\providecommand{\BIBforeignlanguage}[2]{{%
\expandafter\ifx\csname l@#1\endcsname\relax
\typeout{** WARNING: IEEEtran.bst: No hyphenation pattern has been}%
\typeout{** loaded for the language `#1'. Using the pattern for}%
\typeout{** the default language instead.}%
\else
\language=\csname l@#1\endcsname
\fi
#2}}
\providecommand{\BIBdecl}{\relax}
\BIBdecl

\bibitem{Biason2017WCNC}
A.~Biason, S.~Dey, and M.~Zorzi, ``Decentralized transmission policies for
  energy harvesting devices,'' in \emph{arXiv:1610.07284}, accepted for
  presentation at \emph{IEEE Wireless Communications and Networking Conference
  (WCNC)}, Mar. 2017.

\bibitem{Ulukus2015}
S.~Ulukus, A.~Yener, E.~Erkip, O.~Simeone, M.~Zorzi, P.~Grover, and K.~Huang,
  ``Energy harvesting wireless communications: A review of recent advances,''
  \emph{IEEE J. Sel. Areas in Commun.}, vol.~33, no.~3, pp. 360--381, Mar.
  2015.

\bibitem{Erturk2011}
A.~Erturk and D.~J. Inman, \emph{Piezoelectric energy harvesting}.\hskip 1em
  plus 0.5em minus 0.4em\relax John Wiley \& Sons, 2011.

\bibitem{Stordeur1997}
M.~Stordeur and I.~Stark, ``Low power thermoelectric generator-self-sufficient
  energy supply for micro systems,'' in \emph{Proc. IEEE 16th Int. Conf. on
  Thermoelectrics (ICT)}, Aug. 1997, pp. 575--577.

\bibitem{Beeby2006}
S.~P. Beeby, M.~J. Tudor, and N.~M. White, ``Energy harvesting vibration
  sources for microsystems applications,'' \emph{Measurement Science and
  Technology}, vol.~17, no.~12, pp. R175--R195, Oct. 2006.

\bibitem{Raghunathan2005}
V.~Raghunathan, A.~Kansal, J.~Hsu, J.~Friedman, and M.~Srivastava, ``Design
  considerations for solar energy harvesting wireless embedded systems,'' in
  \emph{Proc. IEEE 4th Int. Symp. on Information Processing in Sensor Networks
  (IPSN)}, Apr. 2005, pp. 457--462.

\bibitem{Wang2008}
W.~S. Wang, T.~O'Donnell, N.~Wang, M.~Hayes, B.~O'Flynn, and C.~O'Mathuna,
  ``Design considerations of {sub-mW} indoor light energy harvesting for
  wireless sensor systems,'' \emph{J. Emerg. Technol. Comput. Syst.}, vol.~6,
  no.~2, pp. 6:1--6:26, June 2008.

\bibitem{Lu2015}
X.~Lu, P.~Wang, D.~Niyato, D.~I. Kim, and Z.~Han, ``Wireless networks with {RF}
  energy harvesting: A contemporary survey,'' \emph{IEEE Commun. Surveys
  Tutorials}, vol.~17, no.~2, pp. 757--789, Second Quarter 2015.

\bibitem{Michelusi2013}
N.~Michelusi, K.~Stamatiou, and M.~Zorzi, ``Transmission policies for energy
  harvesting sensors with time-correlated energy supply,'' \emph{IEEE Trans.
  Commun.}, vol.~61, no.~7, pp. 2988--3001, July 2013.

\bibitem{Biason2014}
A.~Biason, D.~Del~Testa, and M.~Zorzi, ``Low-complexity policies for wireless
  sensor networks with two energy harvesting devices,'' in \emph{Proc. IEEE
  13th Annual Mediterranean on Ad Hoc Networking Workshop (MED-HOC-NET)}, June
  2014, pp. 180--187.

\bibitem{Tutuncuoglu2012a}
K.~Tutuncuoglu and A.~Yener, ``Optimum transmission policies for battery
  limited energy harvesting nodes,'' \emph{IEEE Trans. Wireless Commun.},
  vol.~11, no.~3, pp. 1180--1189, Mar. 2012.

\bibitem{Ozel2012b}
O.~Ozel and S.~Ulukus, ``Achieving {AWGN} capacity under stochastic energy
  harvesting,'' \emph{IEEE Trans. Inf. Theory}, vol.~58, no.~10, pp.
  6471--6483, Oct. 2012.

\bibitem{Sharma2010}
V.~Sharma, U.~Mukherji, V.~Joseph, and S.~Gupta, ``Optimal energy management
  policies for energy harvesting sensor nodes,'' \emph{IEEE Trans. Wireless
  Commun.}, vol.~9, no.~4, pp. 1326--1336, Apr. 2010.

\bibitem{Pielli2016}
C.~Pielli, A.~Biason, A.~Zanella, and M.~Zorzi, ``Joint optimization of energy
  efficiency and data compression in {TDMA-based} medium access control for the
  {IoT},'' in \emph{Proc. IEEE Global Communications Conf. (GLOBECOM), IoT-LINK
  Workshop}, Dec. 2016.

\bibitem{Lei2009}
J.~Lei, R.~Yates, and L.~Greenstein, ``A generic model for optimizing
  single-hop transmission policy of replenishable sensors,'' \emph{IEEE Trans.
  Wireless Commun.}, vol.~8, no.~2, pp. 547--551, Feb. 2009.

\bibitem{Ozel2011}
O.~Ozel and S.~Ulukus, ``{AWGN} channel under time-varying amplitude
  constraints with causal information at the transmitter,'' in \emph{Proc. 45th
  Asilomar Conf. on Signals, Systems and Computers (ASILOMAR)}, Nov. 2011, pp.
  373--377.

\bibitem{Smith2013}
J.~R. Smith, \emph{Wirelessly Powered Sensor Networks and Computational
  RFID}.\hskip 1em plus 0.5em minus 0.4em\relax Springer Science \& Business
  Media, 2013.

\bibitem{Biason2015d}
A.~Biason and M.~Zorzi, ``Joint transmission and energy transfer policies for
  energy harvesting devices with finite batteries,'' \emph{IEEE J. Sel. Areas
  in Commun.}, vol.~33, no.~12, pp. 2626--2640, Dec. 2015.

\bibitem{Biason2016b}
------, ``Battery-powered devices in {WPCN}s,'' \emph{IEEE Trans. Commun.},
  Early access 2016.

\bibitem{Blasco2013}
P.~Blasco, D.~Gunduz, and M.~Dohler, ``A learning theoretic approach to energy
  harvesting communication system optimization,'' \emph{IEEE Trans. Wireless
  Commun.}, vol.~12, no.~4, pp. 1872--1882, Apr. 2013.

\bibitem{Alsheikh2015}
M.~A. Alsheikh, D.~T. Hoang, D.~Niyato, H.-P. Tan, and S.~Lin, ``Markov
  decision processes with applications in wireless sensor networks: {A}
  survey,'' \emph{IEEE Commun. Surveys \& Tutorials}, vol.~17, no.~3, pp.
  1239--1267, Third Quarter 2015.

\bibitem{Mohrehkesh2014}
S.~Mohrehkesh and M.~C. Weigle, ``Optimizing energy consumption in {TeraHertz}
  band nanonetworks,'' \emph{IEEE J. Sel. Areas in Commun.}, vol.~32, no.~12,
  pp. 2432--2441, Dec. 2014.

\bibitem{Michelusi2015}
N.~Michelusi and M.~Zorzi, ``Optimal adaptive random multiaccess in energy
  harvesting wireless sensor networks,'' \emph{IEEE Trans. Commun.}, vol.~63,
  no.~4, pp. 1355--1372, Apr. 2015.

\bibitem{Hoang2014}
D.~T. Hoang, D.~Niyato, P.~Wang, and D.~I. Kim, ``Optimal decentralized control
  policy for wireless communication systems with wireless energy transfer
  capability,'' in \emph{Proc. IEEE Int. Conf. Communications (ICC)}, June
  2014, pp. 2835--2840.

\bibitem{Dibangoye2012}
J.~S. Dibangoye, C.~Amato, and A.~Doniec, ``Scaling up decentralized {MDPs}
  through heuristic search,'' in \emph{Proc. Uncertainty in Artificial
  Intelligence (UAI)}, Aug. 2012.

\bibitem{Dibangoye2013}
J.~S. Dibangoye, C.~Amato, A.~Doniec, and F.~Charpillet, ``Producing efficient
  error-bounded solutions for transition independent decentralized {MDPs},'' in
  \emph{Proc. ACM Int. Conf. Autonomous Agents and Multi-agent Systems
  (AAMAS)}, May 2013.

\bibitem{Dibangoye2014}
J.~S. Dibangoye, C.~Amato, O.~Buffet, and F.~Charpillet, ``Optimally solving
  {Dec-POMDPs} as continuous-state {MDPs:} theory and algorithms,''
  \emph{INRIA}, Research Report, no. 8517, Apr. 2014.

\bibitem{Lovejoy1991}
W.~S. Lovejoy, ``Computationally feasible bounds for partially observed
  {Markov} decision processes,'' \emph{Operations research}, vol.~39, no.~1,
  pp. 162--175, Jan.--Feb. 1991.

\bibitem{Zhou2001}
R.~Zhou and E.~A. Hansen, ``An improved grid-based approximation algorithm for
  {POMDPs},'' in \emph{Proc. Int. Joint Conf. on Artificial Intelligence},
  vol.~17, no.~1, Aug 2001, pp. 707--716.

\bibitem{Korf1990}
R.~E. Korf, ``Real-time heuristic search,'' \emph{Artificial intelligence},
  vol.~42, no. 2-3, pp. 189--211, Mar. 1990.

\bibitem{Dechter2003}
R.~Dechter, \emph{Constraint processing}.\hskip 1em plus 0.5em minus
  0.4em\relax Morgan Kaufmann, 2003.

\bibitem{Dechter1997}
------, ``Bucket elimination: a unifying framework for processing hard and soft
  constraints,'' \emph{Constraints}, vol.~2, no.~1, pp. 51--55, Apr. 1997.

\bibitem{Bertsekas2005}
D.~Bertsekas, \emph{Dynamic programming and optimal control}.\hskip 1em plus
  0.5em minus 0.4em\relax Athena Scientific, Belmont, Massachusetts, 2005.

\bibitem{Michelusi2012}
N.~Michelusi, K.~Stamatiou, and M.~Zorzi, ``On optimal transmission policies
  for energy harvesting devices,'' in \emph{Proc. IEEE Information Theory and
  Applications Workshop (ITA)}, Feb. 2012, pp. 249--254.

\bibitem{Wang2015}
X.~Wang, J.~Gong, C.~Hu, S.~Zhou, and Z.~Niu, ``Optimal power allocation on
  discrete energy harvesting model,'' \emph{EURASIP J. Wireless Commun. and
  Networking}, vol. 2015, no.~1, pp. 1--14, Mar. 2015.

\bibitem{Shaviv2015}
D.~Shaviv and A.~{\"O}zg{\"u}r, ``Universally near-optimal online power control
  for energy harvesting nodes,'' in \emph{Proc. IEEE Int. Conf. Communications
  (ICC)}, arXiv:1511.00353, May 2015.

\bibitem{Amato2013}
C.~Amato, G.~Chowdhary, A.~Geramifard, N.~K. {\"U}re, and M.~J. Kochenderfer,
  ``Decentralized control of partially observable {Markov} decision
  processes,'' in \emph{Proc. IEEE 52nd Conf. on Decision and Control}, Dec.
  2013, pp. 2398--2405.

\bibitem{toulbar2}
\BIBentryALTinterwordspacing
``Toulbar2,'' an open source WCSP solver. [Online]. Available:
  \url{https://mulcyber.toulouse.inra.fr/projects/toulbar2/}
\BIBentrySTDinterwordspacing

\end{thebibliography}
\bibliographystyle{IEEEtran}

\end{document}